\documentclass[final,3p,times]{elsarticle}

\usepackage{amssymb}
\usepackage{amsthm}
\usepackage{amsmath}
\usepackage{booktabs}
\usepackage{makecell}
\usepackage{layout}

\usepackage{geometry}
\usepackage{multirow}
\usepackage{tabularx}
\usepackage{algorithm}
\usepackage{algpseudocode}
\usepackage{placeins}
\usepackage{flafter}

\usepackage{setspace}
\usepackage{hyperref}

\usepackage{xcolor}

\journal{Elsevier}

\begin{document}

\begin{frontmatter}

    \title{A Variational Kolosov--Muskhelishvili Network for Elasticity and Fracture}

    \author{Shuwei Zhou\corref{cor1}}
    \ead{shuwei.zhou@rwth-aachen.de}
    \author{Christian Häffner}
    \author{Sophie Stebner}
    \author{Niklas Fehlemann}
    \author{Zhichao Wei}
    \author{Sebastian Münstermann}

    \cortext[cor1]{Corresponding author: Shuwei Zhou.}


    \affiliation{organization={Institute of Metal Forming, RWTH Aachen University},
        city={Aachen},
        postcode={52072},
        country={Germany}}

    \begin{abstract}
        Physics-informed neural networks provide a mesh-free framework for solving partial differential equation-governed problems in solid mechanics. However, most existing formulations in linear elasticity still learn the displacement field directly, which does not explicitly exploit the analytic structure of two-dimensional elasticity and becomes restrictive for fracture problems with crack face discontinuities and crack tip singularities. Moreover, existing Kolosov--Muskhelishvili informed neural network formulations still rely on residual-based loss functions with multiple boundary and interface terms, whereas a variational concept has not yet been established. To address these issues, a variational Kolosov--Muskhelishvili informed neural network framework for two-dimensional linear elastic problems with and without cracks is proposed in this work. The solution is represented by two holomorphic Kolosov--Muskhelishvili potentials and trained through an energy-based loss function derived from the principle of minimum total potential energy. For crack problems, a discontinuous stress potential representation is further adopted to embed the crack face condition and crack tip singularity directly into the solution ansatz. The proposed framework is validated on a series of benchmark problems with and without cracks. The results show that the variational Kolosov--Muskhelishvili informed neural network can accurately predict stress and displacement fields as well as stress intensity factors. Compared with traditional neural network models, the proposed framework provides higher accuracy and a more unified loss construction in the considered cases. Overall, the proposed variational Kolosov--Muskhelishvili informed neural network provides an effective and physically consistent variational framework for two-dimensional linear elastic fracture analysis.
    \end{abstract}




\end{frontmatter}


\section{Introduction}
\label{sec:intr1o}

Partial differential equations (PDEs) are mathematical models of physical laws \cite{Jagtap.2020,Ge.2024,Yue.2026}, which form the foundation of solid mechanics, fluid mechanics, heat transfer, and other engineering disciplines \cite{Liu.2025}. Over the past decades, classical numerical methods such as the finite element method (FEM) \cite{Zhou.2026,Wei.2025,Liu.2020,Skamniotis.2025}, the virtual element method (VEM) \cite{Mora.2026,Leng.2025,Xu.2024}, and the boundary element method (BEM) \cite{Chen.2022b,Gu.2020} have become standard tools for solving such problems. In addition to classical local methods, peridynamics has also become an important framework for fracture and discontinuity problems, because cracks can be represented without the need for spatial derivatives or special crack-tracking techniques~\cite{SILLING2000175,WANG201889}.
Among them, FEM remains the dominant framework because of its solid mathematical foundation and engineering applicability in solid mechanics \cite{Wei.2024,Shen.2023}. Nevertheless, the performance of mesh-based methods still depends strongly on the quality of the discretization \cite{Goswami.2020}. In particular, problems involving stress concentrations \cite{Cao.2025}, geometric discontinuities \cite{Ge.2024}, or crack tip singularities \cite{Nguyen.2015,Zhou.2024b,Roy.2024} typically require local mesh refinement. In addition, evolving cracks \cite{Lu.2019,Wang.2024m}, repeated inverse analyses \cite{Liao.2025b}, and optimization loops \cite{Wei.2022} often demand repeated remeshing and repeated system solves. These issues can increase the computational cost and implementation complexity \cite{Wang.2024o,Pan.2025}.

With the evolution of computational resources, machine learning (ML) methods have shown strong potential for complex mapping problems between input and output \cite{Esterhuizen.2022,Zhou.2024}. The models based on ML have been explored for constitutive modeling \cite{Ibragimova.2022b}, surrogate simulation \cite{Sim.2025,Pagan.2022}, parameter identification \cite{Fehlemann.2023,Kong.2025}, fatigue life prediction \cite{BabuRao.2025,Zhang.2024d,Hu.2024d,Peng.2022}, and accelerated simulation \cite{Wang.2026b}. Compared with traditional numerical methods, ML methods offer higher efficiency in repeated evaluations once the training process is completed \cite{Xu.2023b}. However, traditional machine learning approaches are typically data dependent and often require large amounts of high quality training data \cite{Hu.2024c}. Their black-box character may also limit physical interpretability and reliability when extrapolation is required \cite{Gan.2022b,Zhou.2023b}. These limitations have motivated increasing interest in domain knowledge-informed learning frameworks that incorporate governing physical laws directly into the training process \cite{Yin.2024,Zhu.2025}. 

A representative development along this direction is the physics-informed neural network (PINN), introduced by Raissi et al.~\cite{Raissi.2019}, in which the governing equations and boundary conditions are embedded into the loss function \cite{Cuomo.2022,Gu.2024}. PINN has been successfully applied to solve a variety of problems in fluid flow \cite{Lu.2021}, heat transfer \cite{Cai.2021,Meng.2023}, fatigue life prediction \cite{Salvati.2022,Feng.2025,Zhou.2025,Feng.2024}, and solid mechanics \cite{Weng.2025,Xiong.2025}. PINN libraries such as SciANN \cite{Haghighat.2021} and DeepXDE \cite{Wang.2021b} have been developed to facilitate the construction of PINN formulations. In solid mechanics, PINN formulations are commonly constructed either in strong form, by minimizing the residuals of equilibrium and boundary conditions \cite{Rezaei.2022}, or in variational form, by minimizing an energy functional such as the total potential energy~\cite{Samaniego.2020,Yu.2024c,ZHOU2024117226,Ma.2025}. Compared with strong form formulations, energy-based formulations so-called deep energy methods (DEM) are often more compact \cite{Wang.2022f}. Additionally, energy-based methods can reduce the burden of balancing multiple residual terms \cite{Zheng.2022}. Even so, most existing PINN frameworks in elasticity still approximate the displacement field directly, while stresses and strains are calculated afterward through constitutive and kinematic relations \cite{Wang.2026}. Such a representation does not explicitly exploit the analytic structure of two-dimensional elasticity and becomes particularly restrictive in fracture problems, where displacement discontinuities, free crack faces, and crack-tip singularities must be represented accurately \cite{Gu.2023,Zhang.2026b}. Recent neural network approaches have also been developed for crack propagation, phase-field fracture, and plastic fracture problems, further demonstrating the growing interest in physics-informed learning for fracture mechanics \cite{FENG2025110800,FENG2025118284,Feng.2026}.

For two-dimensional isotropic linear elasticity, the Kolosov--Muskhelishvili (KM) complex-potential formalism provides an exact analytical representation of the stress and displacement field in terms of holomorphic functions \cite{Bock.2015}. This representation offers a natural foundation for physics-informed learning, because the equilibrium and compatibility conditions can be satisfied structurally through the choice of analytic potentials, rather than enforced through additional PDE residual terms \cite{Calafa.2024}. Based on this idea, physics-informed holomorphic neural network (PIHNN) frameworks have recently been developed to learn the two KM potentials directly and reconstruct the corresponding elastic field \cite{Calafa.2025,Hund.2026}. In our previous work, this line of research was further extended to the Kolosov--Muskhelishvili informed neural network (KMINN), in which normalized loss terms were introduced to balance displacement, traction, and interface constraints, and crack enrichment as well as crack propagation criteria were further incorporated for fracture analysis \cite{Wang.2024p,Zhou.2026b}. Despite these advances, existing KM-informed neural formulations still rely mainly on residual-based or multi-term boundary objectives. As a result, the training procedure still depends on the construction and balancing of several loss components, especially for mixed boundary value problems and cracked domains. This means that the variational structure of linear elasticity, in which the exact solution can be characterized as the minimizer of the total potential energy, has not yet been fully exploited within the KM-informed neural setting \cite{Chen.2025}. To the authors' knowledge, a variational KM-informed neural framework that combines holomorphic representation with the principle of minimum total potential energy for two-dimensional fracture problems has not yet been reported.

Motivated by this observation, a variational Kolosov--Muskhelishvili informed neural network (vKMINN) is proposed for two-dimensional linear elastic problems with and without cracks in this study. In the proposed framework, the unknown solution is represented by two holomorphic KM potentials and trained through an energy-based objective derived from the principle of minimum total potential energy. For crack problems, a discontinuous stress potential representation is further adopted so that the crack face condition and the crack tip singularity are embedded directly into the solution ansatz. In this way, the essential analytic structure of plane elasticity and the characteristic fracture field behavior are both incorporated into the model by construction. The resulting formulation remains mesh-free and avoids the explicit use of PDE residual losses in the considered setting. The proposed framework is assessed through several benchmark problems, including elastic domains without cracks, and fracture configurations under mode-I and mixed-mode loading. The numerical results show that the proposed method can accurately predict stress and displacement field and can provide reliable evaluations of stress intensity factors (SIFs). Compared with conventional PINNs based on output of displacement and traditional residual-based KMINN formulations, the proposed approach leads to a more unified loss construction and improved convergence behavior in the cases considered. In this sense, the present work differs mainly in the variational reformulation of KM-informed holomorphic neural approximations through an energy-based loss functional.

The remainder of this paper is organized as follows. Section~\ref{sec:governing_equations_and_complex_variable_formulation} introduces the governing equations, the KM complex-variable representation, and the variational formulation. Section~\ref{sec:methodology} presents the proposed vKMINN framework, including the complex-valued neural architecture, the crack representation, the energy-based loss, and the SIFs evaluation. Section~\ref{sec:case_studies_and_discussions} reports the numerical case studies and discussions. Finally, the main conclusions are summarized in Section~\ref{sec:conclusions}.

\section{Governing Equations and Complex-Variable Formulation}
\label{sec:governing_equations_and_complex_variable_formulation}
In this section, the governing equations, the KM complex-variable representation, and the variational formulation are introduced. Additionally, the SIFs and stress singularity are discussed to provide a physical understanding of the proposed framework.

\subsection{Kolosov-Muskhelishvili Governing Equations}
\label{sec:kolosov-muskhelishvili_governing_equations}

For isotropic, homogeneous, linear elastic materials, the strong form of the 2D equilibrium equations is given by,
\begin{equation}
    \nabla \cdot \boldsymbol{\sigma} + \boldsymbol{f} = \boldsymbol{0}
    \quad \text{in } \Omega,
    \label{eq:equilibrium_strong}
\end{equation}
where $\boldsymbol{\sigma}$ is the Cauchy stress tensor and $\boldsymbol{f}$ denotes the body force (zero in this study).
Hooke's law and the small-strain kinematics can be written as follows,
\begin{equation}
    \displaystyle \boldsymbol{\sigma} = 2\mu\,\boldsymbol{\varepsilon}
    + \lambda \,\operatorname{tr}(\boldsymbol{\varepsilon})\,\boldsymbol{I},
    \qquad
    \displaystyle \boldsymbol{\varepsilon} = \frac{1}{2}\bigl(\nabla \boldsymbol{u}
    + (\nabla \boldsymbol{u})^{\mathsf{T}}\bigr),
    \label{eq:strain_displacement_relation}
\end{equation}
where $\boldsymbol{\varepsilon}$ is the strain tensor, $\boldsymbol{u}$ denotes the displacement field, $\operatorname{tr}(\cdot)$ represents the trace of the tensor, and $\boldsymbol{I}$ is the second-order identity tensor. In addition, the shear modulus and Lamé constants are expressed as,
\begin{equation}
    \mu = \frac{E}{2(1+\nu)}, \qquad
    \lambda = \frac{E\nu}{(1+\nu)(1-2\nu)},
\end{equation}
where $E$ is Young's modulus and $\nu$ is Poisson's ratio. The boundary conditions are given by,
\begin{equation}
    \left\{
    \begin{array}{ll}
        u_i = \bar{u}_i,             & \textrm{on} ~\Gamma_u \quad \textrm{(Dirichlet)}, \\[5pt]
        \sigma_{ij} n_j = \bar{t}_i, & \textrm{on} ~\Gamma_t \quad \textrm{(Neumann)},
    \end{array}
    \label{eq:boundary_conditions}
    \right.
\end{equation}
where $\bar{u}_i$ is the prescribed displacement, $\bar{t}_i$ is the prescribed traction, and $n_j$ is the outward normal vector on the boundary $\Gamma_u$ and $\Gamma_t$. Introducing the complex coordinate $z = x + i y$ and two holomorphic Kolosov--Muskhelishvili potentials $\varphi(z)$ and $\psi(z)$, any plane linear elastic solution can be represented equivalently as \cite{Radaj.2014},
\begin{equation}
    \left\{
    \begin{aligned}
         & \sigma_{xx} + \sigma_{yy} = 4\,\mathrm{Re}\bigl(\varphi'(z)\bigr),                         \\
         & \sigma_{yy} - \sigma_{xx} + 2i\sigma_{xy} = \bar{z}\,\varphi''(z) + \psi'(z),              \\
         & 2\mu\,(u_x + i u_y) = \kappa\,\varphi(z) - z\,\overline{\varphi'(z)} - \overline{\psi(z)},
    \end{aligned}
    \right.
    \label{eq:KM_compact}
\end{equation}
where $\kappa$ is the Kolosov constant, given by $\kappa = (3-\nu)/(1+\nu)$ in plane stress and $\kappa = 3 - 4\nu$ in plane strain, and $\overline{(\cdot)}$ denotes complex conjugation. Thus, all stress and displacement components $\sigma_{xx},\sigma_{yy},\sigma_{xy},u_x,u_y$ can be calculated by two KM potentials $\varphi(z)$ and $\psi(z)$ directly as,
\begin{equation}
    \left\{
    \begin{aligned}
         & \sigma_{xx} (\varphi, \psi)  = \mathrm{Re}\left( 2\varphi'(z) - \bar{z}\,\varphi''(z) - \psi'(z) \right),                                    \\[2pt]
         & \sigma_{yy} (\varphi, \psi)  = \mathrm{Re}\left( 2\varphi'(z) + \bar{z}\,\varphi''(z) + \psi'(z) \right),                                    \\[2pt]
         & \sigma_{xy} (\varphi, \psi)  = \mathrm{Im}\left( \bar{z}\,\varphi''(z) + \psi'(z) \right),                                                   \\[2pt]
         & u_x         (\varphi, \psi) = \frac{1}{2\mu}\,\mathrm{Re}\left( \kappa\,\varphi(z) - z\,\overline{\varphi'(z)} - \overline{\psi(z)} \right), \\[2pt]
         & u_y         (\varphi, \psi) = \frac{1}{2\mu}\,\mathrm{Im}\left( \kappa\,\varphi(z) - z\,\overline{\varphi'(z)} - \overline{\psi(z)} \right). \\
    \end{aligned}
    \right.
    \label{eq:KM_formulations}
\end{equation}

\subsection{Variational Formulation and Energy Methods}
\label{sec:variational_formulation_and_energy_methods}

The displacement and stress field are obtained from the complex potential functions $\varphi(z)$ and $\psi(z)$ using the KM formulations. In this section, the variational formulation and energy functional of the linear elasticity problem are derived. According to the principle of minimum total potential energy, among all admissible displacement field which satisfy the essential boundary conditions, the actual displacement field minimizes the total potential energy.

The total potential energy functional $w(\mathbf{u})$ is defined as,
\begin{equation}
    w(\mathbf{u})
    = w_{\mathrm{int}}(\mathbf{u}) - w_{\mathrm{ext}}(\mathbf{u}).
    \label{eq:total_potential_energy}
\end{equation}
where $w_{\mathrm{int}}(\mathbf{u})$ is the total internal strain energy and $w_{\mathrm{ext}}(\mathbf{u})$ is the potential of external forces. The solution of the DEM problem is given by $\mathbf{u}^* = \mathop{\mathrm{argmin}}_{\mathbf{u}} w(\mathbf{u})$.

For a linear elastic body occupying the two-dimensional domain $\Omega \subset \mathbb{R}^2$ with boundary $\Gamma = \Gamma_u \cup \Gamma_t$, the strain energy density is given by,
\begin{equation}
    w(\boldsymbol\varepsilon)=\frac12 \boldsymbol\varepsilon:\mathbb D:\boldsymbol\varepsilon = \frac12 \boldsymbol\sigma:\mathbb S:\boldsymbol\sigma
    \label{eq:strain_energy_density}
\end{equation}
where $\boldsymbol{\varepsilon}$ is related to the stress field $\boldsymbol{\sigma}$ and the displacement field $\mathbf{u}$ from Eq.~\eqref{eq:strain_displacement_relation}, $\mathbb D$ is usually called the  four-order elastic stiffness tensor. Since the problem is linear elastic, the strain energy density can be equivalently expressed in terms of stresses through the compliance tensor $\mathbb S = \mathbb D^{-1}$ in this research. The total internal strain energy is adopted by stress field $\boldsymbol{\sigma}$ as,
\begin{equation}
    w(\boldsymbol{\sigma})
    = \frac{1}{4\mu}\Big(\sigma_{xx}^{2}+\sigma_{yy}^{2}+\sigma_{zz}^{2}+2\sigma_{xy}^{2}\Big)
    -\frac{\nu}{2E}\big(\sigma_{xx}+\sigma_{yy}+\sigma_{zz}\big)^{2},
    \qquad \sigma_{xz}=\sigma_{yz}=0,
    \label{eq:psi_stress_master}
\end{equation}
where $\sigma_{zz}=0$ for plane stress and $\sigma_{zz}=\nu(\sigma_{xx}+\sigma_{yy})$ for plane strain.
This stress-based form allows the internal energy to be evaluated without explicitly forming $\varepsilon(\nabla u)$ in the energy computation.
In the absence of body forces, the potential of the prescribed tractions $\bar{\mathbf{t}}$ on $\Gamma_t$ corresponds to the work done by external forces, which is expressed as,
\begin{equation}
    w_{\mathrm{ext}}(\mathbf{u})
    = \int_{\Gamma_t} \mathbf{u} \cdot \bar{\mathbf{t}}\,\mathrm{d}\Gamma
    = \int_{\Gamma_t} u_i\,\bar{t}_i\,\mathrm{d}\Gamma.
    \label{eq:external_potential}
\end{equation}

    The equivalence between the strong form and the variational form is classical in linear elasticity. Taking the first variation of the total potential energy functional gives,
    \begin{equation}
        \delta w(\mathbf{u}, \delta\mathbf{u})
        = \int_{\Omega} \boldsymbol{\sigma} : \delta\boldsymbol{\varepsilon}\,\mathrm{d}\Omega
        - \int_{\Gamma_t} \bar{\mathbf{t}} \cdot \delta\mathbf{u}\,\mathrm{d}\Gamma.
        \label{eq:variation_eq}
    \end{equation}
    Using $\delta\boldsymbol{\varepsilon}=\nabla(\delta\mathbf{u})$, the divergence theorem, and $\delta\mathbf{u}=\mathbf{0}$ on $\Gamma_u$,
    \begin{equation}
        \delta w
        = - \int_{\Omega} (\nabla \cdot \boldsymbol{\sigma}) \cdot \delta\mathbf{u}\,\mathrm{d}\Omega
        + \int_{\Gamma_t} (\boldsymbol{\sigma}\mathbf{n} - \bar{\mathbf{t}}) \cdot \delta\mathbf{u}\,\mathrm{d}\Gamma.
    \end{equation}

For $\delta w$ to vanish for all admissible $\delta\mathbf{u}$, the terms in parentheses must vanish independently. This recovers the equilibrium equation $\nabla \cdot \boldsymbol{\sigma} = \mathbf{0}$ in $\Omega$ and the traction condition $\boldsymbol{\sigma}\mathbf{n} = \bar{\mathbf{t}}$ on $\Gamma_t$, confirming the consistency of the variational formulation.

By employing the KM potentials $\varphi(z)$ and $\psi(z)$ introduced in Section~\ref{sec:kolosov-muskhelishvili_governing_equations}, the displacement and stress field in Eqs.~\eqref{eq:total_potential_energy}--\eqref{eq:external_potential} are expressed via the complex relations in Eq.~\eqref{eq:KM_compact}. Consequently, the primary unknowns are replaced by the analytic KM potentials, and the elastic problem is reformulated as the minimization of $w(\mathbf{u}(\varphi,\psi))$ subject to the essential boundary conditions.

\subsection{Stress Intensity Factor and Stress Singularity} \label{sec:sif_and_singularity}
In 2D  linear elastic fracture mechanics (LEFM), the SIFs are fundamental parameters
characterizing the near-tip stress and displacement fields \cite{Yang.2017}. They can be defined as,
\begin{equation}
    K_\mathrm{I} = \lim_{r \to 0} \sqrt{2\pi r}\,\sigma_{yy}(r,0),\qquad
    K_\mathrm{II} = \lim_{r \to 0} \sqrt{2\pi r}\,\sigma_{xy}(r,0),
    \label{eq:sif_i_and_ii}
\end{equation}
where $r$ is the radial distance from the crack tip and $\theta=0$ denotes the
direction ahead of the crack tip.

As the crack tip is approached ($r\to 0$), the stress and displacement fields
exhibit the well-known square-root singular behavior \cite{Radaj.2014},
\begin{equation}
    \left\{
    \begin{aligned}
        \sigma_{ij}(r,\theta) & \sim
        \frac{K_\mathrm{I}}{\sqrt{2\pi r}}\, f_{ij}^{(\mathrm{I})}(\theta)
        + \frac{K_\mathrm{II}}{\sqrt{2\pi r}}\, f_{ij}^{(\mathrm{II})}(\theta), \\
        u_k(r,\theta)         & \sim
        K_\mathrm{I}\sqrt{\frac{r}{2\pi}}\, g_{k}^{(\mathrm{I})}(\theta)
        + K_\mathrm{II}\sqrt{\frac{r}{2\pi}}\, g_{k}^{(\mathrm{II})}(\theta),
    \end{aligned}
    \right.
    \label{eq:asymptotic_behavior}
\end{equation}
where $f_{ij}^{(\mathrm{I})}(\theta)$ and $f_{ij}^{(\mathrm{II})}(\theta)$ are
the universal angular functions for mode I and mode II, respectively, and
$g_{k}^{(\mathrm{I})}(\theta)$ and $g_{k}^{(\mathrm{II})}(\theta)$ denote the
corresponding angular functions for the displacement field. The stress field exhibits a $1/\sqrt{r}$ singularity, whereas the displacement field varies as $\sqrt{r}$ in the vicinity of the crack tip. These asymptotic expressions define the so-called $K$-dominant zone, in which the near-tip fields are fully characterized by the SIFs $K_\mathrm{I}$ and $K_\mathrm{II}$.

\section{Methodology}\label{sec:methodology}

The proposed vKMINN framework combines the KM complex variable formulation with an energy-based variational training strategy. It differs from standard PINN formulations in solid mechanics. The standard PINN learns the displacement field of the input points from the boundary conditions and equilibrium equations or other governing equations. Rather than directly solving the real-valued governing equations, the vKMINN learns the two complex-valued KM potentials from the governing equations and the boundary conditions. The framework and loss function of the vKMINN are introduced in this section.

\subsection{Complex-Valued Neural Network Architecture} \label{sec:complex_valued_neural_network_architecture}

The present neural architecture is built on the physics-informed holomorphic neural network framework of Calafà et al.~\cite{Calafa.2024}. In this framework, complex-valued neural networks (CVNN) with holomorphic activation functions are used to approximate holomorphic potentials, so that the governing equations of two-dimensional linear elasticity can be satisfied by construction through the KM representation. The proposed vKMINN framework extends this approach to the variational setting by replacing the residual-based loss functions with an energy-based objective derived from the principle of minimum total potential energy.

\begin{figure}[!htbp]
    \centering
    \includegraphics[width=0.95\textwidth,height=0.62\textheight,keepaspectratio]{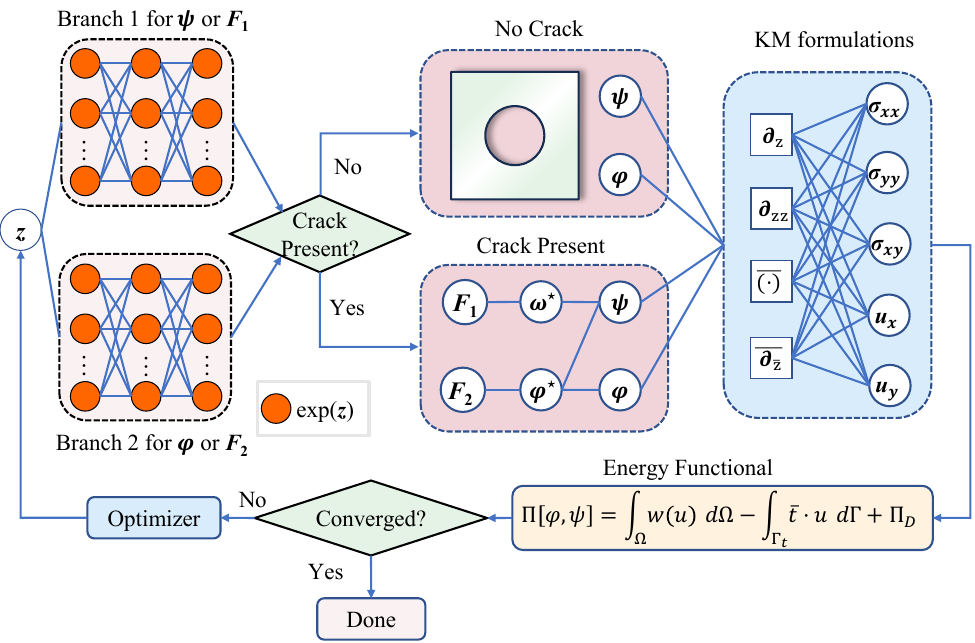}
    \caption{Overall architecture of the proposed vKMINN framework for two-dimensional linear elasticity and fracture problems.}
    \label{fig:vKMINN_architecture}
\end{figure}

The overall architecture of the vKMINN is shown in Fig.~\ref{fig:vKMINN_architecture}. In the vKMINN model, two parallel CVNNs are constructed to approximate the KM potentials $\varphi(z)$ and $\psi(z)$ respectively. The complex coordinate is written as $z=x+\mathrm{i}y\in\Omega\subset\mathbb{C}$. The network outputs two complex-valued functions, denoted by \cite{Lee.2022},
\begin{equation}
    \bigl(\varphi_{\boldsymbol{\vartheta}}(z),\,\psi_{\boldsymbol{\vartheta}}(z)\bigr)=\mathcal{N}_{\boldsymbol{\vartheta}}(z)\in\mathbb{C}^2,
    \label{eq:vKMINN_map_simple}
\end{equation}
where $\mathcal{N}_{\boldsymbol{\vartheta}}=(\mathcal{N}_{{\boldsymbol{\vartheta}}_\varphi},\mathcal{N}_{{\boldsymbol{\vartheta}}_\psi})$ stacks the two subnetworks, and $\boldsymbol{\vartheta}=\{{\boldsymbol{\vartheta}}_\varphi,{\boldsymbol{\vartheta}}_\psi\}$ collects the weights and biases of the two subnetworks. Similar to other ANN frameworks, each subnetwork of the vKMINN is defined layer by layer. The input for the total network is set as,
\begin{equation}
    x^{(0)}(z)=z.
    \label{eq:vKMINN_input_simple}
\end{equation}
For each hidden layer $\ell=1,\dots,L$, a complex affine map is applied first,
\begin{equation}
    y^{(\ell)}(z)=W^{(\ell)}x^{(\ell-1)}(z)+b^{(\ell)},
    \qquad
    W^{(\ell)}\in\mathbb{C}^{n^{(\ell)}_{\mathrm{out}}\times n^{(\ell)}_{\mathrm{in}}},
    \quad
    b^{(\ell)}\in\mathbb{C}^{n^{(\ell)}_{\mathrm{out}}},
    \label{eq:vKMINN_affine_simple}
\end{equation}
and then an activation function is applied component wise,
\begin{equation}
    x^{(\ell)}(z)=\varsigma \left(y^{(\ell)}(z)\right).
    \label{eq:vKMINN_act_simple}
\end{equation}
In this study, an exponential function is used as the activation function to ensure the holomorphicity of the networks following the work \cite{Calafa.2024}. The exponential function is expressed as,
\begin{equation}
    \varsigma(\xi)=\exp(\xi), \qquad \xi\in\mathbb{C}.
    \label{eq:vKMINN_exp_simple}
\end{equation}
A final complex affine map is used to generate the two potentials as follows,

\begin{equation}
    \mathcal{N}_{\boldsymbol{\vartheta}}(z)=
    \begin{bmatrix}
        \varphi_{\boldsymbol{\vartheta}}(z) = W_{\boldsymbol{\vartheta}_\varphi}^{(L+1)}x_{\boldsymbol{\vartheta}_\varphi}^{(L)}(z)+b_{\boldsymbol{\vartheta}_\varphi}^{(L+1)} \\
        \psi_{\boldsymbol{\vartheta}}(z) = W_{\boldsymbol{\vartheta}_\psi}^{(L+1)}x_{\boldsymbol{\vartheta}_\psi}^{(L)}(z)+b_{\boldsymbol{\vartheta}_\psi}^{(L+1)}
    \end{bmatrix}.
    \label{}
\end{equation}

Holomorphicity is required by the KM representation. Therefore, it is enforced by construction in the vKMINN. Since complex affine maps are entire and the exponential function is entire, $\mathcal{N}_{\boldsymbol{\vartheta}}$ is entire as a finite composition of entire functions. Consequently, $\varphi_{\boldsymbol{\vartheta}}$ and $\psi_{\boldsymbol{\vartheta}}$ are holomorphic in $\Omega$ and satisfy the Cauchy--Riemann constraint in the Wirtinger form as given by \cite{Oh.2023},
\begin{equation}
    \frac{\partial \varphi_{\boldsymbol{\vartheta}}}{\partial \bar z}=0,
    \qquad
    \frac{\partial \psi_{\boldsymbol{\vartheta}}}{\partial \bar z}=0,
    \label{eq:CR_wirtinger}
\end{equation}
thereby meeting the analytic regularity required by the KM representation. In addition to holomorphicity, the exponential function offers a convenient differentiation property $(\exp(\xi))'=\exp(\xi)$, which is beneficial for efficiently evaluating the higher-order derivatives entering the governing equations.

Despite the above advantages, the exponential nonlinearity may lead to rapid activation growth and gradient explosion. To stabilize training at early iterations, Calafà et al. \cite{Calafa.2024} proposed an initialization method to perform the layerwise scaling together with a complex-valued variant of He's initialization \cite{He.2015}. For each complex linear layer $\mathcal{L}_\ell:\mathbb{C}^{n_{\mathrm{in}}^{(\ell)}}\to\mathbb{C}^{n_{\mathrm{out}}^{(\ell)}}$, the real and imaginary parts of the weights are independently initialized as given by,
\begin{equation}
    \Re W^{(\ell)}_{ij},\ \Im W^{(\ell)}_{ij}
    \sim \mathcal{G}\!\left(0,\frac{\rho_\ell}{2\,n_{\mathrm{in}}^{(\ell)}}\right),
    \qquad
    b^{(\ell)}=0,
    \label{eq:he_complex_detailed}
\end{equation}
where $\rho_\ell>0$ is a layerwise variance factor. Given a global parameter $\beta>0$ and a prescribed number $M_e$ of pre-stabilizing layers, $\rho_\ell$ is chosen by the data-dependent rule,
\begin{equation}
    \rho_\ell=
    \begin{cases}
        \dfrac{\beta}{\widehat{\mathbb{E}}\!\left[\,|x^{(\ell-1)}|^2\,\right]}, & \ell\le M_e, \\[8pt]
        \beta e^{-\beta},                                                       & \ell> M_e,
    \end{cases}
    \label{eq:rho_rule_detailed}
\end{equation}
where $\widehat{\mathbb{E}}[\cdot]$ denotes the empirical average over an initial mini batch and $|x^{(\ell-1)}|^2$ is computed element-wise and then averaged over neurons and samples. In practice, layers are initialized sequentially: after initializing layer $\ell$, a single feed forward propagation is performed to estimate $\widehat{\mathbb{E}}\!\left[|x^{(\ell)}|^2\right]$, which is then used to set $\rho_{\ell+1}$. In this way, the empirical second moment is kept approximately constant across layers. As additional safeguards, small global variance factors and gradient clipping are also used during the training process.

\subsection{A Discontinuous Stress Potential for Crack Surface} \label{sec:crack_tip_singularity_analysis}

The normal elastic solid mechanics problem can be solved by the vKMINN framework. However, the crack problem is more challenging and requires a special treatment. Referring to the work of Calafà et al. \cite{Calafa.2025}, a Rice-type enriched representation of the crack surface is adopted. Classical LEFM admits a compact complex-variable formulation through the KM potentials. In the present work, the elastic fields are represented by two holomorphic functions $\varphi(z)$ and $\psi(z)$. Following crack representation of previous work \cite{Rice.1968}, it is convenient to introduce the auxiliary KM potential $\omega(z)$, defined as,
\begin{equation}
    \omega(z) := \psi(z) + z\,\varphi'(z).
    \label{eq:omega_def}
\end{equation}
This replacement is standard and will be used throughout.
Consider a straight crack embedded in an infinite medium, located on the segment $\{s\in[-a,a],\, y=0\}$ in the complex plane.
For traction-free crack faces, the boundary condition can be written as a Riemann--Hilbert system for the limiting values of the complex potentials on the crack line \cite{Jiang.2022},
\begin{equation}
    \left\{
    \begin{aligned}
        \varphi^{+}(s) + \omega^{-}(s) & = 0, \\
        \varphi^{-}(s) + \omega^{+}(s) & = 0,
    \end{aligned}
    \right.
    \qquad s\in(-a,a),
    \label{eq:RH_system_varphi_omega}
\end{equation}
where the superscripts $+$ and $-$ denote the limits taken from the upper and lower half-planes, respectively.
By adding the two relations in Eq.~\eqref{eq:RH_system_varphi_omega}, one obtains the homogeneous jump condition,
\begin{equation}
    [\varphi(s)+\omega(s)]^{+} + [\varphi(s)+\omega(s)]^{-} = 0,
    \qquad s\in(-a,a),
    \label{eq:RH_sum}
\end{equation}
which indicates that $\varphi+\omega$ possesses a branch cut along the crack segment. Additionally, the difference $\varphi-\omega$ remains regular across the crack surface.
Therefore, the solution must reproduce the characteristic $\sqrt{r}$ crack tip singularity and discontinuity across the crack surface explicitly \cite{Wang.2023f}.

To encode this behavior explicitly, the geometric characteristic function $\zeta(z)$ is introduced as,
\begin{equation}
    \zeta(z)=
    \begin{cases}
        \sqrt{z^2 - a^2}, & \text{internal crack on } \{x\in[-a,a],\,y=0\}, \\[2mm]
        \sqrt{z},         & \text{edge crack on } \{y=0\},
    \end{cases}
    \label{eq:zeta_def}
\end{equation}
where the branch choice fixes the crack-line cut.
With this notation, Woo's global representation \cite{C.W.WOO.1989} states that there exist two holomorphic functions $F_1(z)$ and $F_2(z)$ such that,
\begin{equation}
    \left\{
    \begin{aligned}
        \varphi(z) & = \zeta(z) F_1(z) + F_2(z),             \\
        \omega(z)  & = \zeta(z) F_1^\star(z) - F_2^\star(z),
    \end{aligned}
    \right.
    \label{eq:modified_rice_varphi_omega}
\end{equation}
where the conjugate transformation is defined by,
\begin{equation}
    F^\star(z) := \overline{F(\overline{z})}.
    \label{eq:star_def}
\end{equation}
This representation enforces both the crack-line discontinuity and the $\sqrt{r}$ singularity through $\zeta(z)$, while $F_1(z)$ and $F_2(z)$ remain bounded and holomorphic even on the crack line.

In the proposed vKMINN framework, the regular holomorphic functions $F_1(z)$ and $F_2(z)$ are approximated by two CVNNs with entire activation functions, denoted by $\mathcal{N}_{F_1}(z,\boldsymbol{\vartheta}_{F_1})$ and $\mathcal{N}_{F_2}(z,\boldsymbol{\vartheta}_{F_2})$.
Accordingly, the prediction ansatz is chosen as \cite{Zhou.2026b},
\begin{equation}
    \varphi_{\mathrm{pred}}(z)= \zeta(z)\,\mathcal{N}_{F_1}(z,{\boldsymbol{\vartheta}}_{F_1}) + \mathcal{N}_{F_2}(z,{\boldsymbol{\vartheta}}_{F_2}),
    \label{eq:vKM_varphi_ansatz}
\end{equation}
and,
\begin{equation}
    \omega_{\mathrm{pred}}(z)= \zeta(z)\,\mathcal{N}_{F_1}^\star(z,{\boldsymbol{\vartheta}}_{F_1}) - \mathcal{N}_{F_2}^\star(z,{\boldsymbol{\vartheta}}_{F_2}).
    \label{eq:vKM_omega_ansatz}
\end{equation}
Note that the $\mathcal{N}^\star(z,\vartheta):=\overline{\mathcal{N}(\overline z,\vartheta)}$. Finally, the second KM potential is recovered through Eq.~\eqref{eq:omega_def} as,
\begin{equation}
    \psi_{\mathrm{pred}}(z)=\omega_{\mathrm{pred}}(z) - z\,\varphi_{\mathrm{pred}}'(z).
    \label{eq:vKM_psi_recover}
\end{equation}
For traction-free straight cracks, the above construction satisfies the crack face condition by design, and therefore no explicit traction-free loss term on the crack faces is required.

\subsection{Energy-based Loss and Numerical Integration}\label{sec:energy_based_loss_and_numerical_integration}

In the present work, the stresses are not treated as independent fields but are generated by the two KM complex potentials $\varphi(z)$ and $\psi(z)$ through Eq.~\eqref{eq:KM_compact}. The $\varphi(z)$ and $\psi(z)$ are approximated by neural networks training process. The mechanical fields $\boldsymbol{\sigma}(\varphi,\psi)$ and $\boldsymbol{u}(\varphi,\psi)$ are obtained through the analytic relations by mathematical identity in Eq.~\eqref{eq:KM_compact}. The network parameters are determined by minimizing a total potential energy functional. In this setting, the elastic solution is obtained as a stationary point of the potential energy over the admissible space.

For linear elasticity without body forces, the energy-based loss is defined from the total potential energy functional as,
\begin{equation}
    \Pi[\varphi,\psi]=\int_{\Omega} w(\boldsymbol{\sigma}(\varphi,\psi))\,\mathrm{d}\Omega-\int_{\Gamma_t}\bar{\boldsymbol{t}}\cdot \boldsymbol{u}(\varphi,\psi)\,\mathrm{d}\Gamma+\alpha_u\int_{\Gamma_u}\bigl|\boldsymbol{u}(\varphi,\psi)-\bar{\boldsymbol{u}}\bigr|^2\,\mathrm{d}\Gamma,
    \label{eq:potential_energy_functional}
\end{equation}
where $\alpha_u>0$ controls the strength of the Dirichlet enforcement. It should be mentioned that the first and second terms are used to calculate internal and external potential energy, respectively. The third term is used to enforce the Dirichlet boundary condition on $\Gamma_u$. $\alpha_u=1000$ is used after repeated tuning in the present work. The desired elastic solution corresponds to a stationary point of $\Pi[\varphi,\psi]$ in the admissible function space. In practice, the neural network parameters are determined by minimizing the discrete counterpart of Eq.~\eqref{eq:potential_energy_functional}.

In standard FEM, the domain integral in Eq.~\eqref{eq:potential_energy_functional} is evaluated by quadrature over the mesh nodes. Instead, the domain integral is evaluated by Monte Carlo quadrature over sampling points in the physical domain. Let  $\{z^{(m)}\}_{m=1}^{N_{\Omega}} \subset \Omega$ be a set of points that are drawn (quasi-)uniformly in $\Omega$. Denoting by $|\Omega|$ the area of the domain, the approximate expression is,
\begin{equation}
    \int_{\Omega}
    w\bigl(\boldsymbol{\sigma}(\varphi,\psi)\bigr)~\mathrm{d}\Omega \approx |\Omega|\frac{1}{N_\Omega}\sum_{m=1}^{N_\Omega}w\bigl(\boldsymbol{\sigma}(\varphi,\psi)\bigr)\big|_{z=z^{(m)}}.
    \label{eq:MC_domain_potential}
\end{equation}
For simply connected geometries, $|\Omega|$ can be computed analytically. For geometrically complex domains like a cracked plate, $|\Omega|$ is estimated by an outer bounding box and a Monte Carlo area estimator. A large number of points is drawn uniformly in the bounding box, and the area is recovered as the bounding-box area multiplied by the fraction of points that fall inside $\Omega$. In cracked configurations, a small core region around each crack tip is excluded from the sampling to avoid the singular behavior of the elastic fields in the energy density. Additionally, the sampling grid is refined around the crack tip to ensure the accuracy of the integration.

The boundary integrals in Eq.~\eqref{eq:potential_energy_functional} are approximated in the same mesh-free method. The boundary $\partial\Omega$ is represented as a finite collection of analytic curves $\Gamma_k$, each with a known arc length $|\Gamma_k|$ and a prescribed boundary condition. For each curve $\Gamma_k$, a set of $N_{\Gamma_k}$ sampling points $\{z_k^{(q)}\}_{q=1}^{N_{\Gamma_k}}$ is generated and distributed quasi-uniformly with respect to the arc-length parameter. The Dirichlet penalty term on $\Gamma_u$ is then approximated by,
\begin{equation}
    \int_{\Gamma_u} \bigl\| \boldsymbol{u}(\varphi,\psi) - \bar{\boldsymbol{u}} \bigr\|^{2} \,\mathrm{d}\Gamma \approx \sum_{\Gamma_k \subset \Gamma_u} |\Gamma_k|\, \frac{1}{N_{\Gamma_k}} \sum_{q=1}^{N_{\Gamma_k}} \bigl\| \boldsymbol{u}(\varphi,\psi) - \bar{\boldsymbol{u}} \bigr\|^{2} \Big|_{z = z_k^{(q)}}.
    \label{eq:MC_boundary_u}
\end{equation}

Similarly, the external work term on $\Gamma_t$ is approximated by,
\begin{equation}
    \int_{\Gamma_t} \bar{\boldsymbol{t}}\cdot \boldsymbol{u}(\varphi,\psi)\,\mathrm{d}\Gamma \approx \sum_{\Gamma_k \subset \Gamma_t} |\Gamma_k|\frac{1}{N_{\Gamma_k}}\sum_{q=1}^{N_{\Gamma_k}}\bar{\boldsymbol{t}}\cdot \boldsymbol{u}(\varphi,\psi)\big|_{z=z_k^{(q)}}.
    \label{eq:MC_boundary_t}
\end{equation}

Eqs. \eqref{eq:MC_domain_potential}--\eqref{eq:MC_boundary_t} provide a consistent Monte Carlo discretization of the continuous functional Eq.~\eqref{eq:potential_energy_functional}. This discretization defines the energy-based loss used to train the vKMINN. The resulting scheme remains fully mesh-free, since the geometry enters only through the analytic boundary description and an indicator function used to classify interior points, while both domain and boundary integrals are evaluated by stochastic quadrature over scattered sampling points.

\subsection{Stress Intensity Factor Evaluation}\label{sec:sif_evaluation}

The near-tip fields are fully characterized by the mode~I and mode~II SIFs, $K_I$ and $K_{II}$, in LEFM theory. In the present work, $K_I$ and $K_{II}$ are extracted from the vKMINN solution using an interaction-integral method, which is path-independent for linear elasticity in the absence of body forces and provides a robust, global measure of the crack tip driving force \cite{Su.2023,Wang.2024g}.

For a two-dimensional isotropic linear-elastic body, the Rice $J$-integral is written as \cite{Zhang.2025b},
\begin{equation}
    J[\boldsymbol{u}]
    \;=\;
    \oint_{\Theta}
    \Big(
    w\,n_x
    -
    (\boldsymbol{\sigma} \boldsymbol{n})\cdot \partial_x \boldsymbol{u}
    \Big)\, \mathrm{d}s,
    \label{eq:j_def}
\end{equation}
where $\Theta$ is a closed contour enclosing the crack tip, $\boldsymbol{n}=(n_x,n_y)$ is the outward unit normal, $w=\tfrac12\,\boldsymbol{\sigma}:\boldsymbol{\varepsilon}$ is the strain energy density, and $\partial_x \boldsymbol{u}=\partial \boldsymbol{u}/\partial x$ is taken in a local crack tip coordinate system (with $x$ tangent to the crack). In practice, $\Theta$ is chosen as a circle centered at the tip and the integral is evaluated by numerical quadrature using the smooth KMINN fields.

For isotropic elasticity, $J$ is related to the SIFs by,
\begin{equation}
    J = \frac{1}{E'}\left(K_I^2+K_{II}^2\right),
    \qquad
    E'=
    \begin{cases}
        E,                  & \text{plane stress}, \\[2pt]
        \dfrac{E}{1-\nu^2}, & \text{plane strain}.
    \end{cases}
    \label{eq:j_sif}
\end{equation}

Substituting the superimposed fields $\boldsymbol{u}^{\mathrm{tot}} = \boldsymbol{u} + \boldsymbol{u}^{\mathrm{aux}}$ and $\boldsymbol{\sigma}^{\mathrm{tot}} = \boldsymbol{\sigma} + \boldsymbol{\sigma}^{\mathrm{aux}}$ into the definition of the $J$-integral in Eq.~\eqref{eq:j_def}, the total energy release rate is expanded as
\begin{equation}
    J[\boldsymbol{u} + \boldsymbol{u}^{\mathrm{aux}}]
    = \oint_{\Theta} \left[ \frac{1}{2}\left(\sigma_{ij}+\sigma_{ij}^{\mathrm{aux}}\right)\left(\varepsilon_{ij}+\varepsilon_{ij}^{\mathrm{aux}}\right) n_x
        - \left(t_i+t_i^{\mathrm{aux}}\right)\,\frac{\partial (u_i+u_i^{\mathrm{aux}})}{\partial x} \right] \,\mathrm{d}s.
\end{equation}
By grouping terms associated exclusively with the actual state $(\boldsymbol{u}, \boldsymbol{\sigma})$ and the auxiliary state $(\boldsymbol{u}^{\mathrm{aux}}, \boldsymbol{\sigma}^{\mathrm{aux}})$, the integral can be decomposed into self-energy terms and interaction terms,
\begin{equation}
    J[\boldsymbol{u} + \boldsymbol{u}^{\mathrm{aux}}]
    = J[\boldsymbol{u}] + J[\boldsymbol{u}^{\mathrm{aux}}] + I[\boldsymbol{u}, \boldsymbol{u}^{\mathrm{aux}}].
\end{equation}
The interaction integral $I$ collects the cross-terms,
\begin{equation}
    I[\boldsymbol{u}, \boldsymbol{u}^{\mathrm{aux}}]
    = \oint_{\Theta}
    \left[
        \left(\boldsymbol{\sigma}:\boldsymbol{\varepsilon}^{\mathrm{aux}}\right) n_x
        - \boldsymbol{t} \cdot \frac{\partial \boldsymbol{u}^{\mathrm{aux}}}{\partial x}
        - \boldsymbol{t}^{\mathrm{aux}} \cdot \frac{\partial \boldsymbol{u}}{\partial x}
        \right] \,\mathrm{d}s.
    \label{eq:interaction_integral_explicit}
\end{equation}
In the LEFM context, Eq.~\eqref{eq:j_sif} implies,
\begin{equation}
    J[\boldsymbol{u}+\boldsymbol{u}^{\mathrm{aux}}] = \frac{1}{E'} \left[ (K_I + K_I^{\mathrm{aux}})^2 + (K_{II} + K_{II}^{\mathrm{aux}})^2 \right].
\end{equation}
Expanding the quadratic terms yields,
\begin{equation}
    J[\boldsymbol{u}+\boldsymbol{u}^{\mathrm{aux}}]
    = \underbrace{\frac{K_I^2 + K_{II}^2}{E'}}_{J[\boldsymbol{u}]}
    + \underbrace{\frac{(K_I^{\mathrm{aux}})^2 + (K_{II}^{\mathrm{aux}})^2}{E'}}_{J[\boldsymbol{u}^{\mathrm{aux}}]}
    + \frac{2}{E'} \left(K_I K_I^{\mathrm{aux}} + K_{II} K_{II}^{\mathrm{aux}}\right).
\end{equation}
Comparing this expansion with the integral decomposition defines the mutual energy release rate relation,
\begin{equation}
    I[\boldsymbol{u},\boldsymbol{u}^{\mathrm{aux}}]
    =
    \frac{2}{E'}
    \left(
    K_I K_I^{\mathrm{aux}} + K_{II} K_{II}^{\mathrm{aux}}
    \right),
    \label{eq:i_sif_rewrite}
\end{equation}
where $(K_I^{\mathrm{aux}},K_{II}^{\mathrm{aux}})$ are the auxiliary SIFs associated with $(\boldsymbol{u}^{\mathrm{aux}},\boldsymbol{\sigma}^{\mathrm{aux}})$. Two auxiliary choices are used: (i) a unit mode~I field with $(K_I^{\mathrm{aux}},K_{II}^{\mathrm{aux}})=(1,0)$, and (ii) a unit mode~II field with $(K_I^{\mathrm{aux}},K_{II}^{\mathrm{aux}})=(0,1)$.
Denoting the corresponding interaction integrals by $I^{(I)}$ and $I^{(II)}$, Eq.~\eqref{eq:i_sif_rewrite} yields the SIFs directly,
\begin{equation}
    K_I = \frac{E'}{2}\, I^{(I)},
    \qquad
    K_{II} = \frac{E'}{2}\, I^{(II)}.
    \label{eq:sif_from_i_rewrite}
\end{equation}

In theory, $I$ is independent of the chosen contour $\Theta$ for linear elasticity without body forces. In the present work, the crack face condition and the crack tip singularity are satisfied by construction via the discontinuous stress potential method. The only practical limitation is that the contour should not be too close to the external boundary; otherwise, any reasonable contour radius can be used to obtain identical results. In the present work, 500 integral points are used for the contour integration in the interaction-integral method. The advantage of imposing this hard constraint via discontinuous stress potentials will be compared with the Williams enrichment used in the KMINN method.

\section{Case Studies and Discussions}
\label{sec:case_studies_and_discussions}
In this section, a series of numerical case studies are conducted to validate the proposed vKMINN model. For problems without crack, the stress and displacement components predicted by vKMINN are compared against analytical solutions and the FEM results. In addition, three cracked cases are considered to assess the effectiveness of vKMINN in evaluating stress and displacement fields as well as SIFs, using FEM or analytical solutions as references. All case studies were conducted on Python 3.13.5 and PyTorch 2.7 using NVIDIA H100 GPUs. Adam optimizer is employed for all models. No learning-rate decay is used, and the training is terminated after a prescribed number of iterations. To evaluate the accuracy of the vKMINN framework, the absolute error, relative $L_2$ error, and coefficient of determination $R^2$ are employed to compare the stress and displacement fields between the vKMINN and PINN. For a generic field quantity $\phi$ (stress or displacement component), these quantities are defined as,
\begin{equation}
    \left\{
    \begin{aligned}
         & e^{abs}_{\phi} = | \phi^{pred} - \phi^{ref} |,                                                                                                                                        \\
         & e^{L_2}_{\phi} = \frac{\| \phi^{pred} - \phi^{ref} \|_2}{\| \phi^{ref} \|_2} = \frac{\sqrt{\sum_{i=1}^{N} (\phi_i^{pred} - \phi_i^{ref})^2}}{\sqrt{\sum_{i=1}^{N} (\phi_i^{ref})^2}}, \\
         & R^2_{\phi} = 1 - \frac{\sum_{i=1}^{N} (\phi_i^{pred} - \phi_i^{ref})^2}{\sum_{i=1}^{N} (\phi_i^{ref} - \bar{\phi}^{ref})^2},
    \end{aligned}
    \right.
\end{equation}
where $\phi^{pred}$ and $\phi^{ref}$ denote the predicted and reference field values, respectively, $\bar{\phi}^{ref} = \frac{1}{N}\sum_{i=1}^{N}\phi_i^{ref}$ is the mean of the reference values, and $N$ is the number of evaluation points. In the first case, the reference solution is obtained from the analytical stress and displacement equations. In the other cases, the reference solution is obtained from the FEM simulations. For the FEM reference solutions, the FEM meshes contain more than 5000 elements and were refined until the reference quantities stabilized.

\subsection{Crack-free Problems}
\label{sec:crack-free_problems}
\begin{figure}[!htbp]
    \centering
    \includegraphics[width=0.92\textwidth,height=0.58\textheight,keepaspectratio]{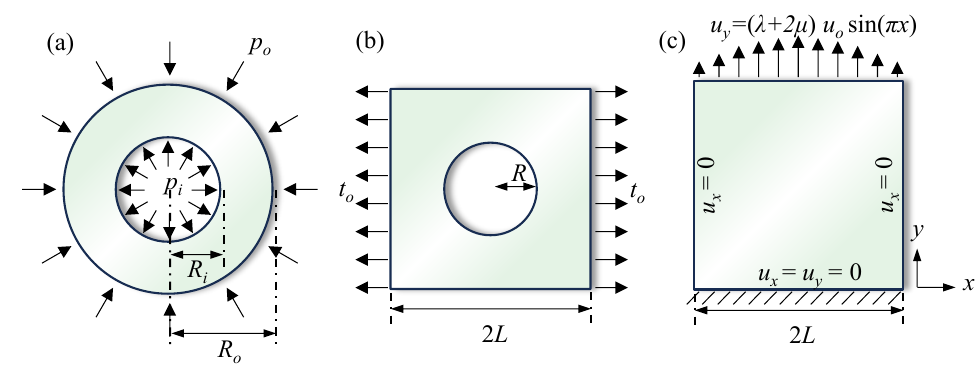}
    \caption{Geometries, loading conditions, and boundary conditions of the benchmark problems without cracks, (a) circular tube under pressure loading, (b) square plate with a hole under tension loading, (c) non-uniform tension of a plate.}
    \label{fig:no-crack_problems}
\end{figure}

In this section, three numerical cases are presented to demonstrate the applicability of the vKMINN framework to problems without cracks, as illustrated in Fig.~\ref{fig:no-crack_problems}. The first case is a circular tube under pressure loading, which is used to verify the accuracy of stress and displacement distributions in an axisymmetric geometry. The second case is a square plate with a hole under tension loading, which is a classical benchmark for evaluating the ability to resolve high stress concentrations near the hole boundary. The third case is a rectangular plate subjected to a sinusoidal loading on the top edge, while the bottom edge is fixed. Moreover, the elastic parameters are set to $\lambda = 1$ and $\mu = 1$ MPa under plane strain conditions for methodological study.

The first two benchmarks are compared only with a conventional PINN. These two cases are mainly governed by pure traction boundary conditions and therefore are not sufficiently informative for clearly distinguishing the proposed variational training strategy from an existing holomorphic boundary-residual framework. In these settings, both PIHNN and vKMINN already benefit from the KM representation, so the difference between residual-based boundary fitting and energy-based training is less pronounced. By contrast, PIHNN is additionally included in the third case because the mixed boundary conditions provide a more suitable setting for isolating the benefit of the proposed variational formulation from that of the holomorphic representation itself.

\subsubsection{Circular Tube Under Pressure Loading}
\label{sec:circular_tube_under_pressure_loading}

The first case considers a circular tube subjected to pressure loading. This benchmark problem has an analytical solution and is used to validate the accuracy of vKMINN in evaluating the stress and displacement fields. The geometry and boundary conditions are shown in Fig.~\ref{fig:no-crack_problems}(a). The inner radius is $R_i = 0.5$ mm, and the outer radius is $R_o = 1$ mm. An internal pressure of $p_i=5$ MPa is applied on the inner surface, and an external pressure of $p_o=10$ MPa is applied on the outer surface. In this case, pressure is taken as positive in compression.

The analytical solutions for the stress and displacement fields are given by,

\begin{equation}
    \left\{
    \begin{aligned}
        \sigma_{xx}(x, y) & = \frac{R_i^2 p_i - R_o^2 p_o}{R_o^2 - R_i^2} - \frac{R_i^2 R_o^2 (p_i - p_o)}{R_o^2 - R_i^2} \cdot \frac{x^2 - y^2}{(x^2 + y^2)^2},                              \\
        \sigma_{yy}(x, y) & = \frac{R_i^2 p_i - R_o^2 p_o}{R_o^2 - R_i^2} + \frac{R_i^2 R_o^2 (p_i - p_o)}{R_o^2 - R_i^2} \cdot \frac{x^2 - y^2}{(x^2 + y^2)^2},                              \\
        \sigma_{xy}(x, y) & = - \frac{2 R_i^2 R_o^2 (p_i - p_o)}{R_o^2 - R_i^2} \cdot \frac{xy}{(x^2 + y^2)^2},                                                                               \\
        u_x(x, y)         & = \frac{1+\nu}{E} \cdot x \cdot \left[ (1-2\nu) \frac{R_i^2 p_i - R_o^2 p_o}{R_o^2 - R_i^2} + \frac{R_i^2 R_o^2 (p_i - p_o)}{(R_o^2 - R_i^2)(x^2 + y^2)} \right], \\
        u_y(x, y)         & = \frac{1+\nu}{E} \cdot y \cdot \left[ (1-2\nu) \frac{R_i^2 p_i - R_o^2 p_o}{R_o^2 - R_i^2} + \frac{R_i^2 R_o^2 (p_i - p_o)}{(R_o^2 - R_i^2)(x^2 + y^2)} \right].
    \end{aligned}
    \right.
\end{equation}

The problem is symmetric with respect to both coordinate axes. Therefore, only one quadrant of the domain is used for evaluation. Symmetry boundary conditions are imposed on the two symmetry lines. On the boundary $x=0$, the following conditions are enforced,
\begin{equation}
    \left\{
    \begin{aligned}
        u_x(0, y) = 0, \\
        \sigma_{xy}(0, y) = 0.
    \end{aligned}
    \label{eq:symmetry_boundary_conditions1}
    \right.
\end{equation}
On the boundary $y=0$, the conditions are
\begin{equation}
    \left\{
    \begin{aligned}
        u_y(x, 0) = 0, \\
        \sigma_{xy}(x, 0) = 0.
    \end{aligned}
    \label{eq:symmetry_boundary_conditions2}
    \right.
\end{equation}

The number of training points is set to 1000, and the number of training iterations is set to 2000. Different combinations of hidden layers, neurons, and learning rates are examined to guide the network design and training settings. The relative $L_2$ errors of the stress and displacement components are summarized in Table~\ref{tbl:circular_tube_under_pressure_loading_comparison}. The best performance is obtained by architecture (B), which uses three hidden layers with 20 neurons per layer (20-20-20) and a learning rate of $1\times 10^{-2}$. For deeper or wider networks, smaller learning rates are also tested to improve training stability. Overall, several architectures achieve satisfactory accuracy when reasonable network sizes and learning rates are adopted.

\begin{table}[!htbp]
    \caption{Comparison of the $L_2$ error of stress and displacement fields between different architectures and learning rates for the vKMINN framework.}
    \label{tbl:circular_tube_under_pressure_loading_comparison}
    \centering
    \begin{tabular}{llllllll}
        \toprule
        Case & NN architecture & Learning rate        & $e^{L_2}_{\sigma_{xx}}$ & $e^{L_2}_{\sigma_{yy}}$ & $e^{L_2}_{\sigma_{xy}}$ & $e^{L_2}_{u_x}$         & $e^{L_2}_{u_y}$         \\
        \midrule
        (A)  & 10-10-10        & 1 $\times$ 10$^{-2}$ & 5.09 $\times$ 10$^{-3}$ & 3.32 $\times$ 10$^{-3}$ & 1.18 $\times$ 10$^{-2}$ & 5.29 $\times$ 10$^{-3}$ & 4.79 $\times$ 10$^{-3}$ \\
        (B)  & 20-20-20        & 1 $\times$ 10$^{-2}$ & 1.29 $\times$ 10$^{-3}$ & 1.29 $\times$ 10$^{-3}$ & 4.95 $\times$ 10$^{-3}$ & 1.54 $\times$ 10$^{-3}$ & 1.46 $\times$ 10$^{-3}$ \\
        (C)  & 20-20-20        & 5 $\times$ 10$^{-3}$ & 4.51 $\times$ 10$^{-3}$ & 6.07 $\times$ 10$^{-3}$ & 1.14 $\times$ 10$^{-2}$ & 1.55 $\times$ 10$^{-2}$ & 1.31 $\times$ 10$^{-2}$ \\
        (D)  & 30-30-30        & 1 $\times$ 10$^{-2}$ & 1.58 $\times$ 10$^{-3}$ & 2.40 $\times$ 10$^{-3}$ & 6.07 $\times$ 10$^{-3}$ & 2.38 $\times$ 10$^{-3}$ & 3.92 $\times$ 10$^{-3}$ \\
        (E)  & 30-30-30        & 5 $\times$ 10$^{-3}$ & 1.53 $\times$ 10$^{-3}$ & 1.62 $\times$ 10$^{-3}$ & 5.26 $\times$ 10$^{-3}$ & 2.96 $\times$ 10$^{-3}$ & 2.48 $\times$ 10$^{-3}$ \\
        (F)  & 20-20-20-20     & 1 $\times$ 10$^{-2}$ & 6.39 $\times$ 10$^{-3}$ & 4.14 $\times$ 10$^{-3}$ & 1.44 $\times$ 10$^{-2}$ & 1.74 $\times$ 10$^{-2}$ & 1.98 $\times$ 10$^{-2}$ \\
        (G)  & 20-20-20-20     & 5 $\times$ 10$^{-3}$ & 3.14 $\times$ 10$^{-3}$ & 1.96 $\times$ 10$^{-3}$ & 8.70 $\times$ 10$^{-3}$ & 6.37 $\times$ 10$^{-3}$ & 3.68 $\times$ 10$^{-3}$ \\
        (H)  & 20-20-20-20-20  & 1 $\times$ 10$^{-2}$ & 9.45 $\times$ 10$^{-3}$ & 4.81 $\times$ 10$^{-3}$ & 2.04 $\times$ 10$^{-2}$ & 1.99 $\times$ 10$^{-2}$ & 1.51 $\times$ 10$^{-2}$ \\
        \bottomrule
    \end{tabular}
\end{table}

A traditional PINN is used as a baseline for comparison. The PINN model follows the framework proposed by SciANN \cite{Haghighat.2021}. The PINN uses real-valued activation functions, in contrast to the complex-valued activation functions used in vKMINN. Therefore, the Swish function is employed as the activation function, which is defined as,
\begin{equation}
    f(x) = x\cdot\frac{1}{1 + e^{-x}}.
\end{equation}

For a fair comparison, the PINN uses the same network size as vKMINN, with three hidden layers and 20 neurons per layer. The learning rate is also set to $1\times 10^{-2}$. To obtain results comparable to those of vKMINN, the number of PINN training iterations is set to 5000.

To provide a consolidated measure of the stress field, the displacement fields and the von Mises equivalent stress are used to evaluate the performance of the two frameworks. The von Mises equivalent stress reflects the stress gradient across the geometry of the case and also highlights the peak stress near the inner boundary. It is defined as,
\begin{equation}
    \sigma_{vm} = \sqrt{\sigma_{xx}^2 + \sigma_{yy}^2 - \sigma_{xx}\sigma_{yy} + 3\sigma_{xy}^2}.
\end{equation}

\begin{figure}[!htbp]
    \centering
    \includegraphics[width=\textwidth,height=0.72\textheight,keepaspectratio]{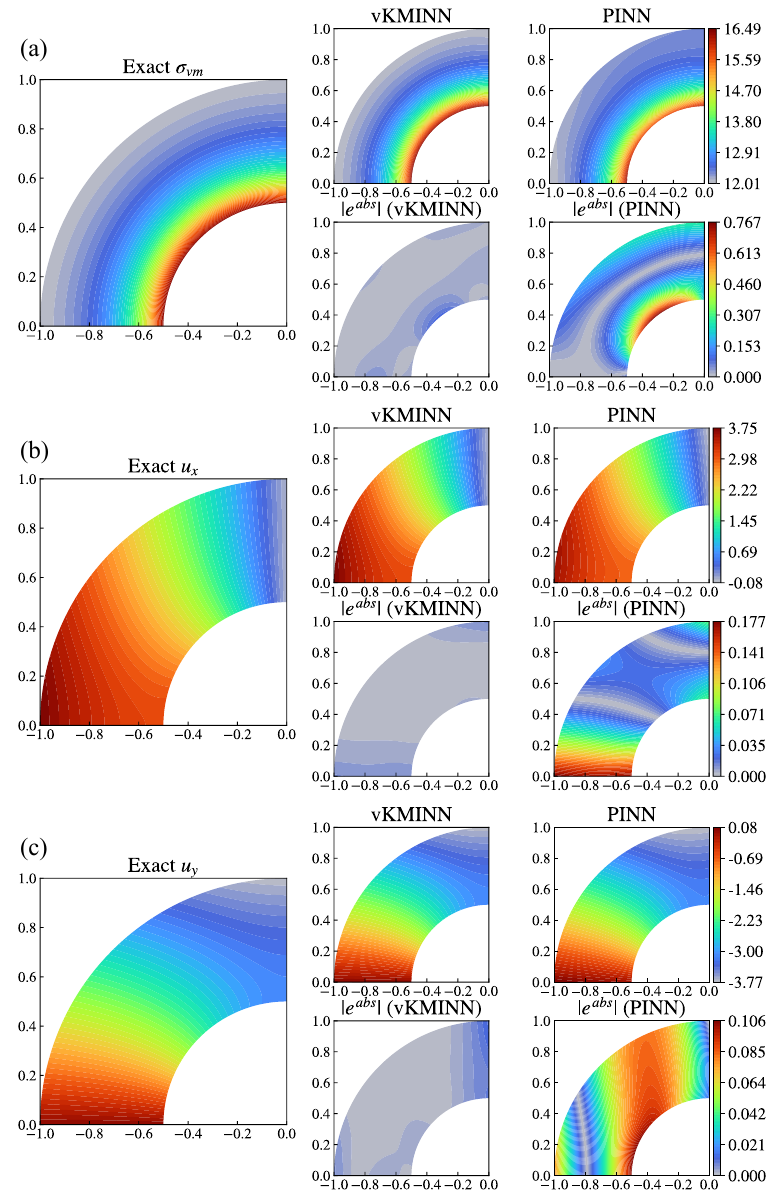}
    \caption{Comparison of von Mises stress and displacement fields, and their absolute errors between vKMINN and PINN for the circular tube under pressure loading: (a) von Mises stress $\sigma_{vm}$, (b) horizontal displacement $u_x$, and (c) vertical displacement $u_y$.}
    \label{fig:circular_tube_under_pressure_loading_comparison}
\end{figure}

The comparison results are shown in Fig.~\ref{fig:circular_tube_under_pressure_loading_comparison}. The vKMINN framework accurately captures the stress and displacement fields with analytical solution, showing smaller absolute errors in the displacement fields and the von Mises equivalent stress than the PINN framework. Overall, vKMINN reproduces the analytical solution with high accuracy and yields consistently lower errors than the PINN baseline for this pressure loaded tube problem. This case also provides guidance for selecting the network architecture and learning rate for subsequent examples.

\subsubsection{Square Plate with a Hole Under Tension Loading}
\label{sec:square_plate_with_a_hole_under_tension_loading}

The second case considers a square plate with a hole subjected to tension loading. A clear stress concentration field is expected near the hole boundary. Specifically, the plate size is $2L=5$ mm, and the hole radius is $R=1$ mm. The plate is subjected to a uniform tensile loading of $t_o=1$ MPa. The detailed geometry is shown in Fig.~\ref{fig:no-crack_problems}(b).

Because the analytical solution corresponds to the Kirsch problem for an infinite plate, the reference solution in this case is obtained from FEM simulations. The FEM model is constructed using CPE4R elements, and the element size is 0.04 mm. Similarly, only the upper-left quadrant is used for evaluation. Symmetry boundary conditions are imposed on the bottom and left boundaries according to Eqs.~\eqref{eq:symmetry_boundary_conditions1} and \eqref{eq:symmetry_boundary_conditions2}. The training settings of vKMINN and PINN are the same as those used in the previous case.

\begin{figure}[!htbp]
    \centering
    \includegraphics[width=\textwidth,height=0.72\textheight,keepaspectratio]{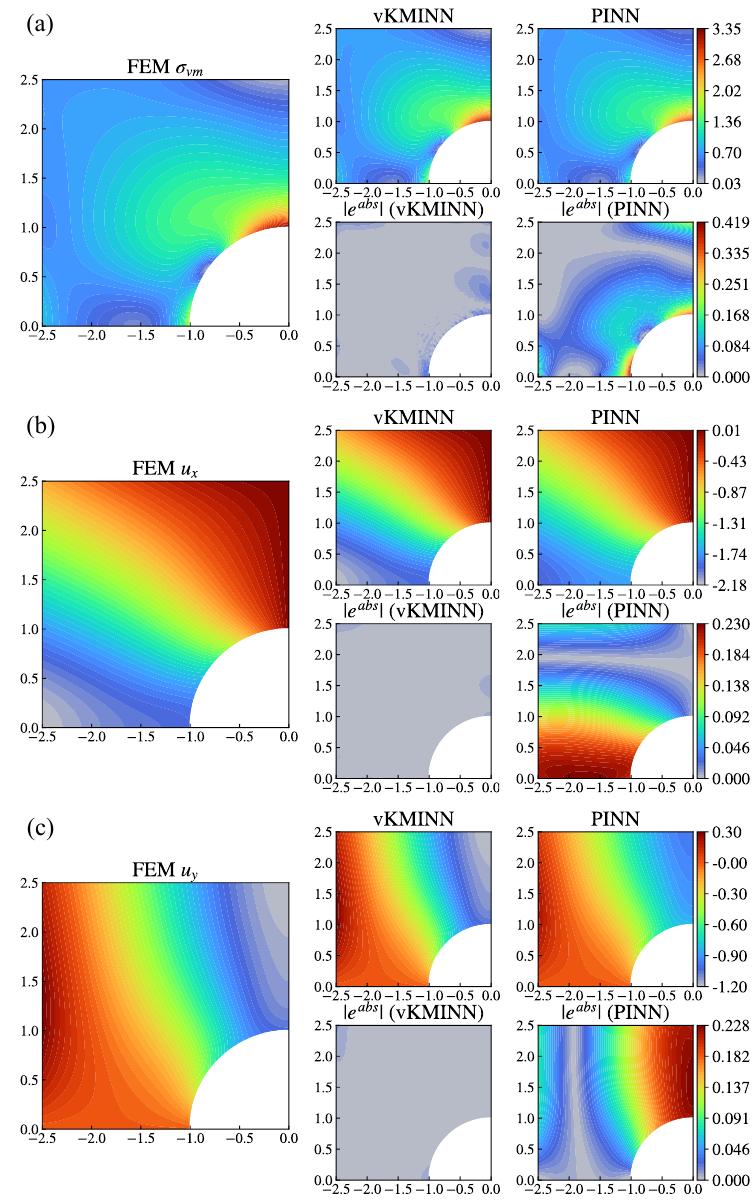}
    \caption{Comparison of the von Mises stress and displacement fields, and their absolute errors between vKMINN and PINN for the square plate with a hole under tension: (a) von Mises stress $\sigma_{vm}$, (b) horizontal displacement $u_x$, and (c) vertical displacement $u_y$.}
    \label{fig:hole_case}
\end{figure}

The $L_2$ errors of the stress and displacement fields for vKMINN and PINN are summarized in Table~\ref{tbl:square_plate_with_a_hole_under_tension_loading_comparison}. The corresponding comparison results are shown in Fig.~\ref{fig:hole_case}. The errors of the vKMINN framework are on the order of $10^{-2}$ and are consistently lower than those of the PINN framework.

\begin{figure}[!htbp]
    \centering
    \includegraphics[width=0.9\textwidth,height=0.64\textheight,keepaspectratio]{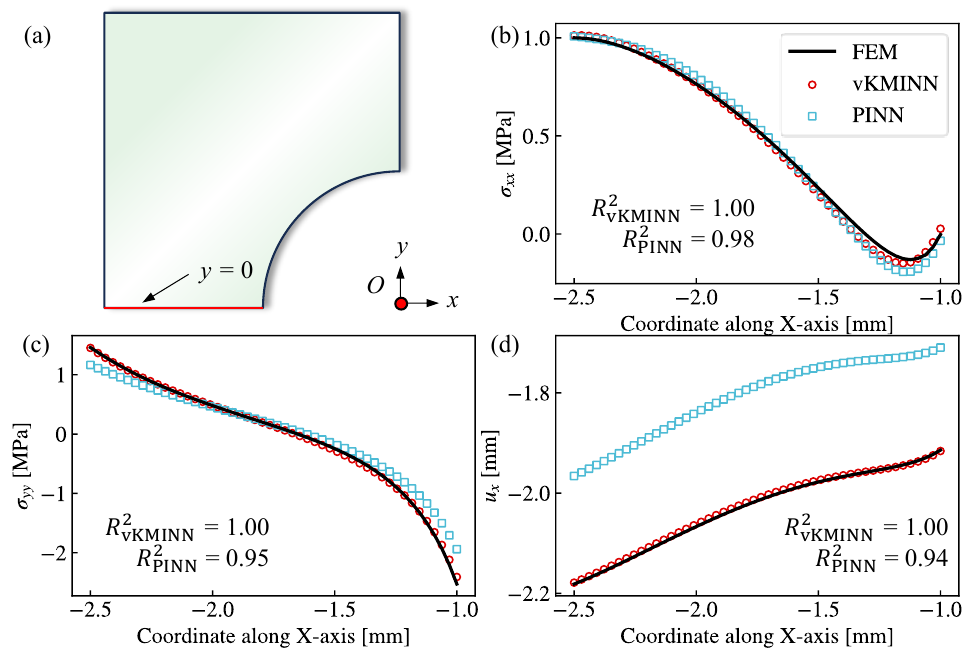}
    \caption{Comparison of the stress and displacement components along the selected path for the square plate with a hole under tension: (a) selected evaluation path, (b) stress component $\sigma_{xx}$, (c) stress component $\sigma_{yy}$, and (d) displacement component $u_x$.}
    \label{fig:case2_comparison}
\end{figure}

\begin{table}[!htbp]
    \centering
    \caption{Comparison of the $L_2$ error of stress and displacement fields between the vKMINN and PINN for the square plate with a hole case.}
    \label{tbl:square_plate_with_a_hole_under_tension_loading_comparison}
    \begin{tabular}{llllll}
        \toprule
               & $e^{L_2}_{\sigma_{xx}}$ & $e^{L_2}_{\sigma_{yy}}$ & $e^{L_2}_{\sigma_{xy}}$ & $e^{L_2}_{u_x}$         & $e^{L_2}_{u_y}$         \\
        \midrule
        vKMINN & 8.29 $\times$ 10$^{-3}$ & 2.61 $\times$ 10$^{-2}$ & 2.30 $\times$ 10$^{-2}$ & 2.05 $\times$ 10$^{-3}$ & 3.30 $\times$ 10$^{-3}$ \\
        PINN   & 6.04 $\times$ 10$^{-2}$ & 2.09 $\times$ 10$^{-1}$ & 1.29 $\times$ 10$^{-1}$ & 9.50 $\times$ 10$^{-2}$ & 1.89 $\times$ 10$^{-1}$ \\
        \bottomrule
    \end{tabular}
\end{table}
To further demonstrate the accuracy of the vKMINN framework, the numerical results along the red line shown in Fig.~\ref{fig:case2_comparison}(a) are extracted from FEM, PINN, and vKMINN. Due to the symmetry boundary conditions, $\sigma_{xx}$, $\sigma_{yy}$, and $u_x$ are used to evaluate the accuracy of the three frameworks. As shown in Fig.~\ref{fig:case2_comparison}(b), (c), and (d), the vKMINN framework consistently provides a better fit to the FEM results than the PINN framework for both the stress and displacement components. In particular, for the displacement component $u_x$, the PINN framework shows a significant deviation from the FEM results. The coefficient of determination $R^2$ is further used to evaluate the predictive capability of the two frameworks.

The $R^2$ value measures the goodness of fit between the predicted and reference values. It ranges from 0 to 1, and a value closer to 1 indicates a better fit. As illustrated in Figs.~\ref{fig:case2_comparison}(b), (c), and (d), the $R^2$ values of the vKMINN framework are all greater than 0.999, which are much higher than those of the PINN framework. These results indicate that the vKMINN framework can accurately predict the stress and displacement fields near the hole boundary.

\subsubsection{Non-uniform Tension of a Plate}
\label{sec:non-uniform_tension_of_plate}

In this case, the boundary conditions are modified to be non-uniform tension loading. The bottom boundary is employed to apply a fixed displacement of $u_x = u_y = 0$. The left and right boundaries are prescribed normal displacement of $u_x = 0$. The top boundary is subjected to a sinusoidal displacement, given by,
\begin{equation}
    u_y = u_0 (\lambda + 2 \mu) \sin(\pi x).
\end{equation}

The geometry of the plate is $2L = 1$ mm, the $\lambda = 1$, $\mu = 1$ MPa, and $u_0 = 0.1$ mm. Thus, the maximum displacement is $u_0 (\lambda + 2 \mu) = 0.3$ mm on the middle of the top boundary.

To further verify the capability of the vKMINN framework, a PIHNN framework is built for comparison under the non-uniform tension loading. Similar to vKMINN, the PIHNN framework searches for the $\varphi$ and $\psi$ potentials from boundary data. In this pure displacement boundary condition case, the loss function is defined as,
\begin{equation}
    L = [\|\boldsymbol{u}^{pred} - \boldsymbol{u}^{ref}\|^2_2]
\end{equation}
where $\boldsymbol{u}^{pred}$ is the calculated displacement on the boundary from the holomorphic network, $\boldsymbol{u}^{ref}$ is the prescribed boundary condition displacement. The training settings of vKMINN and PINN are the same as those used in the previous case. In addition, the PIHNN framework uses the same training hyperparameters as vKMINN.

\begin{figure}[!htbp]
    \centering
    \includegraphics[width=\textwidth,height=0.72\textheight,keepaspectratio]{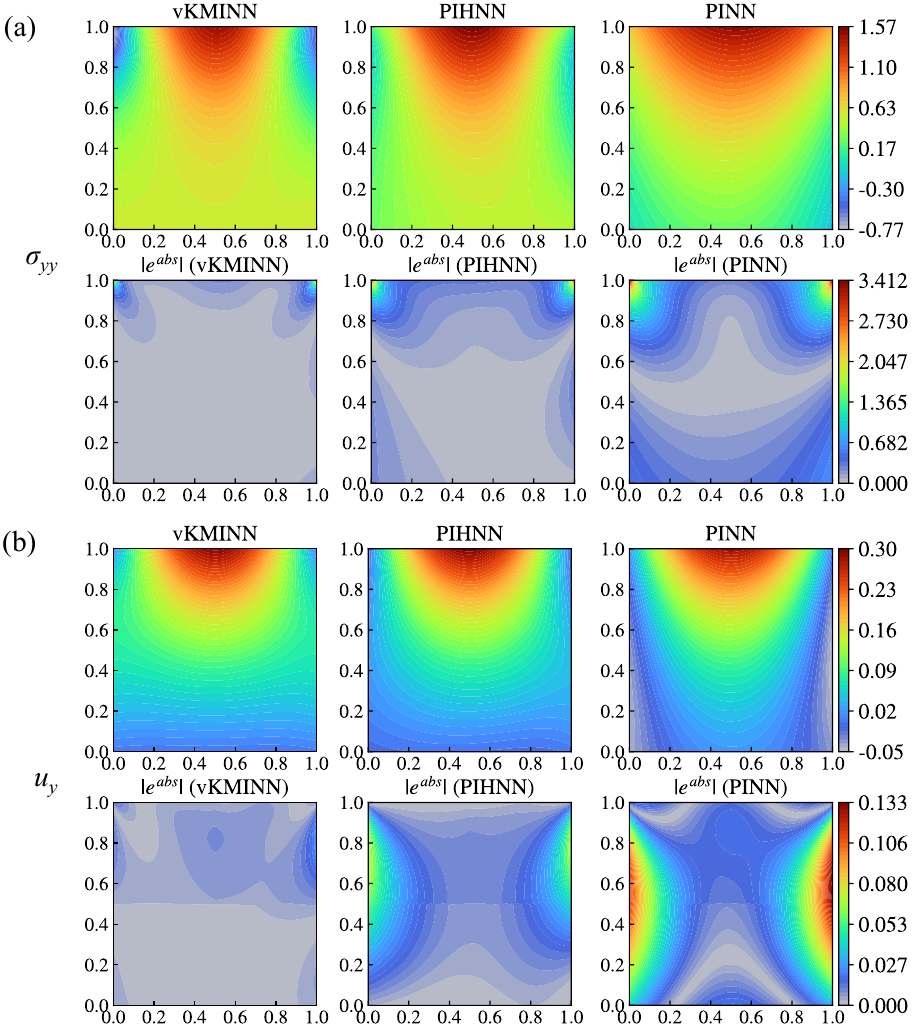}
    \caption{Comparison of $\sigma_{yy}$ and $u_y$ fields and the corresponding absolute errors for vKMINN, PIHNN, and PINN in the plate under non-uniform tension loading: (a) stress component $\sigma_{yy}$ and its absolute error, and (b) displacement component $u_y$ and its absolute error.}
    \label{fig:non-uniform_tension_loading_comparison}
\end{figure}

After training, the results of the displacement component $u_y$ and the stress component $\sigma_{yy}$ from vKMINN, PIHNN, and PINN are compared with the FEM results in Fig.~\ref{fig:non-uniform_tension_loading_comparison}. The results show that the vKMINN framework achieves better prediction accuracy than the PIHNN and PINN frameworks. 

To further examine the difference between the boundary collocation holomorphic formulation and the proposed variational formulation, the training efficiency and stabilized relative $L_2$ errors of vKMINN, PIHNN, and PINN are summarized in Table~\ref{tab:sinload_l2_stable}. The results show that PIHNN reaches a stabilized solution with fewer iterations and a shorter wall-clock time than vKMINN in the present crack-free case, indicating its high training efficiency under pure displacement boundary conditions. However, the final relative $L_2$ errors of vKMINN remain significantly lower than those of PIHNN and PINN for both the $\sigma_{yy}$ and $u_y$ components. This suggests that, although the proposed variational formulation does not necessarily reduce the training time in this case, it provides a clear advantage in the final prediction accuracy for displacement boundary conditions.
\begin{table}[!htbp]
    \centering
    \caption{Comparison of training efficiency and stabilized relative $L_2$ errors for the $\sigma_{yy}$ and $u_y$ components.}
    \label{tab:sinload_l2_stable}
    \begin{tabular}{lccccc}
        \hline
        Model  & Stable iteration & Stable time (s) & Iter/s & $e^{L_2}_{\sigma_{yy}}$ & $e^{L_2}_{u_y}$ \\
        \hline
        vKMINN & 5200             & 53.02           & 98.08  & 0.0763                  & 0.0157          \\
        PIHNN  & 4400             & 22.22           & 198.00 & 0.2117                  & 0.0963          \\
        PINN   & 4400             & 53.84           & 81.72  & 0.3548                  & 0.1824          \\
        \hline
    \end{tabular}
\end{table}

\begin{figure}[!htbp]
    \centering
    \includegraphics[width=\textwidth,height=0.66\textheight,keepaspectratio]{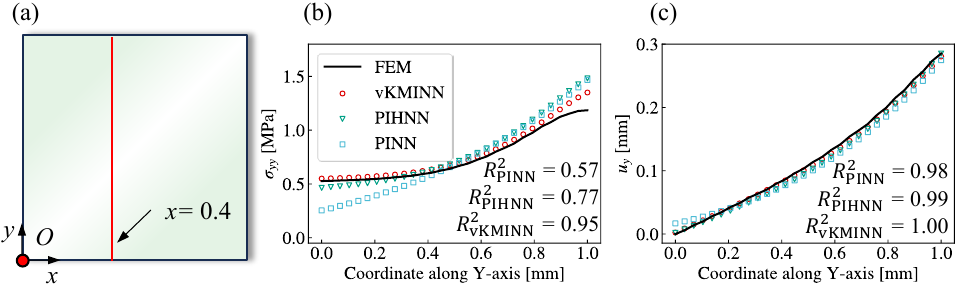}
    \caption{Comparison of the predicted stress and displacement components along the selected vertical path for the plate under non-uniform tension loading: (a) selected evaluation path at $x=0.4$, (b) stress component $\sigma_{yy}$, and (c) displacement component $u_y$.}
    \label{fig:non-uniform_tension_loading_path_comparison}
\end{figure}

The red path in Fig.~\ref{fig:non-uniform_tension_loading_path_comparison}(a) is selected to further evaluate the accuracy of the three frameworks. Figs.~\ref{fig:non-uniform_tension_loading_path_comparison}(b) and (c) show the comparison of the displacement component $u_y$ and the stress component $\sigma_{yy}$ along the red path. The results show that all three frameworks predict the displacement field with reasonable accuracy, although vKMINN performs slightly better than PIHNN and PINN. However, for the stress component $\sigma_{yy}$ along the red path, vKMINN shows significantly higher accuracy than the other two frameworks.

\subsection{Crack Problems}
\label{sec:crack_problems}
In this section, three cases are utilized to evaluate the performance of the proposed vKMINN framework for embedding crack problems. The first case is a single edge notch tension (SENT) problem. The second case is a double edge notch tension (DENT) problem. The third case is an oblique center crack tension (OCCT) problem with non-uniform loading. For the SENT case, the comparison is mainly performed against KMINN. The reason is that both methods are based on the KM representation and crack enrichment. Therefore, the comparison can focus directly on the effect of the loss construction, namely residual-based boundary enforcement in KMINN and the variational energy-based formulation in vKMINN. For the OCCT case, discontinuity-embedded neural network (DENN) is further included as an additional baseline. This is because the OCCT problem is a mixed-mode fracture benchmark, and DENN provides a representative variational neural framework for comparison. In this way, the proposed vKMINN can be assessed against both a KM-based residual framework and an existing energy-based neural method.

\subsubsection{Single Edge Notch Tension}

\begin{figure}[!htbp]
    \centering
    \includegraphics[width=0.9\textwidth,height=0.66\textheight,keepaspectratio]{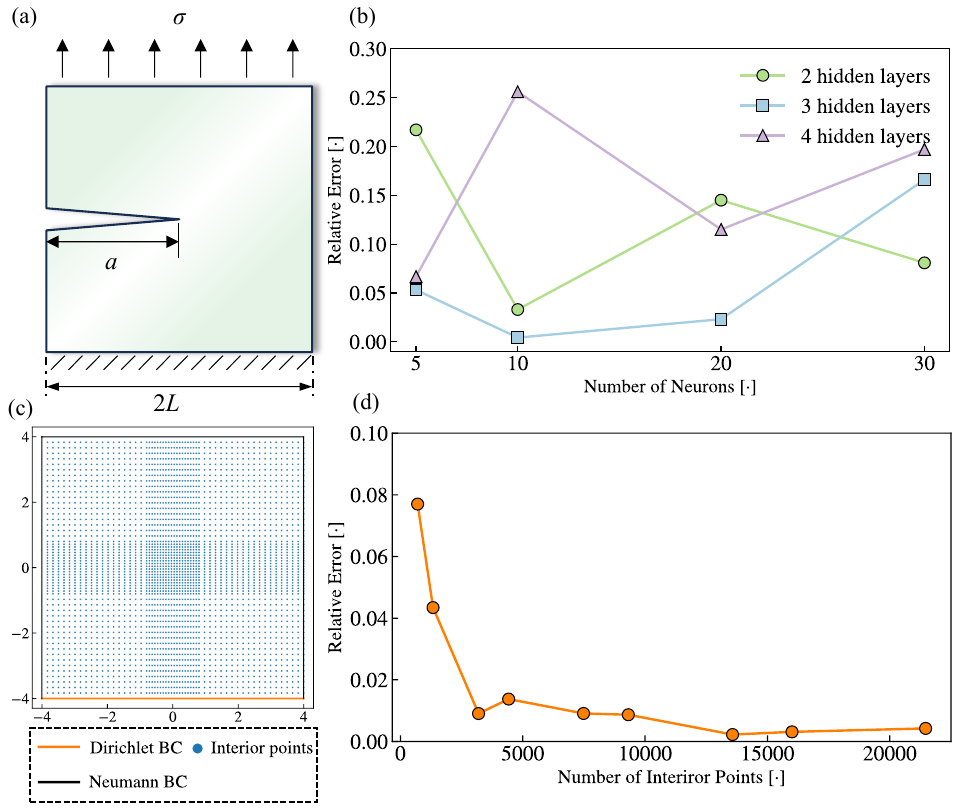}
    \caption{Analysis of the SENT case: (a) geometry and loading conditions of the SENT specimen, (b) relative error of the crack opening displacement (COD) for different neural network architectures, (c) distribution of sampling points and applied boundary conditions, and (d) relative COD error for different numbers of interior points.}
    \label{fig:SENT_COD_comparison}
\end{figure}

The first crack case investigates the performance of the vKMINN framework for a pure mode I crack problem in a SENT specimen. As shown in Fig.~\ref{fig:SENT_COD_comparison}(a), the plate geometry is $2L = 2h = 8$ mm. A uniform tensile load of $\sigma_0 = 1$ MPa is applied on the external boundary, and the lower boundary is fixed. The material properties are $E = 210$ GPa and $\nu = 0.3$.

First, different vKMINN architectures are examined to evaluate the model accuracy and identify the optimal network configuration, as shown in Fig.~\ref{fig:SENT_COD_comparison}(b). The crack opening displacement (COD) at the crack mouth is used to assess the predictive capability of the model for different architectures. The analytical solution for COD is given by~\cite{Tada.1973},

\begin{equation}
    \mathrm{COD}
    = \dfrac{4\sigma_0 a_0}{E^\prime} \,
    \dfrac{\,1.46 + 3.42(1-\cos\beta)\,}{\cos^{2}\beta},
    \qquad
    \beta = \dfrac{\pi a_0}{2L},
    \qquad
    0.2 \leq a_0/L \leq 0.7.
\end{equation}
By comparing the predicted and analytical COD values, the optimal network architecture is selected as 10-10-10 in Fig.~\ref{fig:SENT_COD_comparison}(b).

Furthermore, the influence of the number of sampling points in the domain is investigated to ensure optimal model performance. Wu et al.~\cite{Wu.2023} showed that residual-based adaptive sampling can significantly improve the performance and accuracy of PINN-type models. However, the proposed framework in this study is based on minimization of the total potential energy functional rather than the residual of the governing equations. Therefore, residual-based adaptive sampling is not directly applicable here. Instead, a structured grid with refined sampling points near the crack tip is used to improve the overall performance of the vKMINN framework~\cite{Zhao.2025}, as illustrated in Fig.~\ref{fig:SENT_COD_comparison}(c). In each coordinate direction, the domain is divided into three bands, which together form a $3\times3$ tensor product partition. The refinement is specified by the configuration $(n_a,n_b,n_a,n_b)$, where $n_a$ and $n_b$ denote the numbers of subdivisions used in the outer and crack tip bands, respectively, along both the $x$- and $y$-directions. Candidate grid points located outside the physical domain and inside the excluded crack tip core of radius $r_{\min}$ are removed before evaluating the strain-energy functional. The influence of the quadrature refinement level is further examined in Fig.~\ref{fig:SENT_COD_comparison}(d) and Table~\ref{tbl:sent_sampling_sensitivity}. As the refinement level increases, the COD relative error decreases markedly at the beginning, confirming that sufficient local resolution near the crack tip is essential for capturing the singular field accurately. For finer refinements, however, the error does not decrease monotonically. This indicates that the overall prediction accuracy is influenced not only by the sampling density, but also by the neural network optimization. Among the tested configurations, $(40,40,40,40)$ shows the lowest COD relative error of $0.22\%$, corresponding to 14400 interior points, and is therefore adopted in the subsequent crack analyses as a compromise between accuracy and computational cost.

\begin{table}[!htbp]
    \centering
    \caption{Sensitivity of the SENT prediction to the crack-tip refined sampling density.}
    \label{tbl:sent_sampling_sensitivity}
    \begin{tabular}{ccc}
        \toprule
        Refinement configuration & Points number $N_{\mathrm{est}}$ & COD relative error \\
        \midrule
        $(10,10,10,10)$                              & 900                              & 0.0770             \\
        $(10,20,10,20)$                              & 1600                             & 0.0430             \\
        $(20,20,20,20)$                              & 3600                             & 0.0091             \\
        $(20,30,20,30)$                              & 4900                             & 0.0138             \\
        $(30,30,30,30)$                              & 8100                             & 0.0091             \\
        $(30,40,30,40)$                              & 10000                            & 0.0087             \\
        $(40,40,40,40)$                              & 14400                            & 0.0022             \\
        $(40,50,40,50)$                              & 16900                            & 0.0031             \\
        $(50,50,50,50)$                              & 22500                            & 0.0042             \\
        \bottomrule
    \end{tabular}
\end{table}

\begin{figure}[!htbp]
    \centering
    \includegraphics[width=\textwidth,height=0.72\textheight,keepaspectratio]{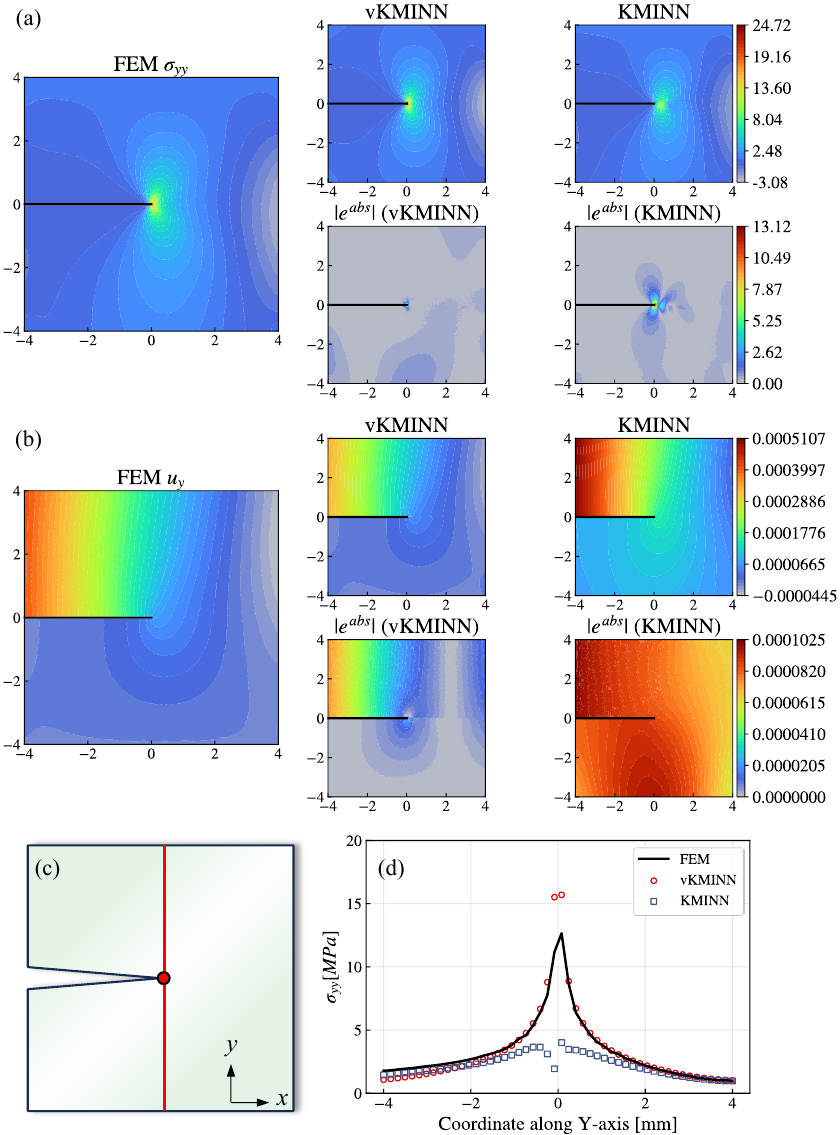}
    \caption{Comparison of FEM, vKMINN, and KMINN results for the SENT case: (a) stress component $\sigma_{yy}$ and its absolute error, (b) displacement component $u_y$ and its absolute error, (c) selected evaluation path, and (d) comparison of $\sigma_{yy}$ along the selected vertical path crossing the crack tip.}
    \label{fig:stress_and_displacement_comparison_between_vKMINN_and_KMINN_for_SENT_case}
\end{figure}

The KMINN framework with Williams enrichment from our previous work \cite{Zhou.2026b} is further used as a reference to assess the accuracy of the proposed vKMINN framework for the mixed boundary value problem in the SENT case. The same training settings are adopted for both frameworks to ensure a fair comparison. As shown in Fig.~\ref{fig:stress_and_displacement_comparison_between_vKMINN_and_KMINN_for_SENT_case}(a) and (b), the vKMINN predictions agree well with those of KMINN for both the stress and displacement fields, and both are in close agreement with the FEM results. Although KMINN introduces a pre-normalized boundary loss to improve the balance among different boundary condition terms, it still requires the treatment of several loss components, including displacement, traction, crack surface, and interface losses. By contrast, the vKMINN framework is formulated through minimization of the total potential energy with an additional displacement penalty term, which leads to a more unified loss functional. Therefore, the vKMINN framework is not only easier to implement but also more efficient in practical training.

To further compare the crack tip behavior, the $\sigma_{yy}$ distribution along a path crossing the crack tip is examined in Fig.~\ref{fig:stress_and_displacement_comparison_between_vKMINN_and_KMINN_for_SENT_case}(c). The results show that both frameworks predict the far field stress with high accuracy. However, near the crack tip, the KMINN results show a more noticeable deviation from the FEM reference. In contrast, the vKMINN framework captures a sharper stress variation and reproduces the crack tip singular behavior more clearly. It should also be noted that the FEM solution exhibits a relatively weaker singular profile near the crack tip, mainly because the singular field is numerically regularized by the finite element discretization and post processing procedure. As a result, the vKMINN prediction appears more singular than the FEM contour near the crack tip.

More specifically, the SIF is used to provide a more comprehensive evaluation of the crack tip stress singularity. The mode I SIF of the SENT problem is given by~\cite{Tada.1973},

\begin{equation}
    K_I = \sqrt{\pi a_0} \cdot \sigma_0 \cdot F(\dfrac{a_0}{L}), \label{eq:sif_for_SENT_case}
\end{equation}
where $F(\dfrac{a_0}{L})$ is the geometric factor for the SENT case, expressed as,

\begin{equation}
    \displaystyle
    F(\dfrac{a_0}{L})=\sqrt{\dfrac{2L}{\pi a_0}\,\tan\!\left(\dfrac{\pi a_0}{2L}\right)}\cdot\dfrac{
        0.752 + 2.02\left(\dfrac{a_0}{L}\right)
        + 0.37\left(1-\sin\!\left(\dfrac{\pi a_0}{2L}\right)\right)^{3}
    }{\cos\!\left(\dfrac{\pi a_0}{2L}\right)}.
    \label{eq:geometric_factor_for_SENT_case}
\end{equation}

\begin{figure}[!htbp]
    \centering
    \includegraphics[width=0.9\textwidth,height=0.58\textheight,keepaspectratio]{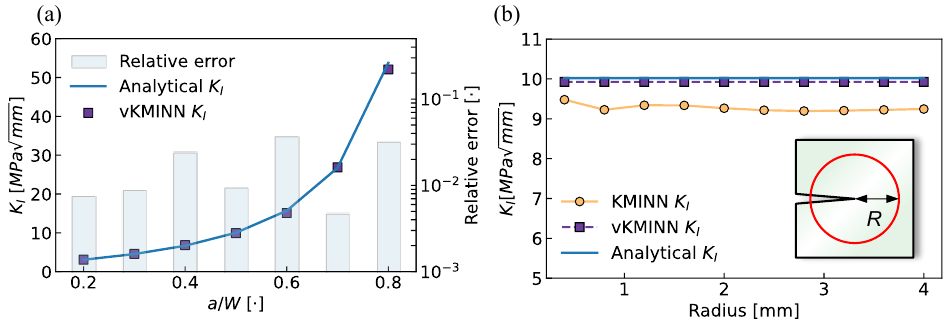}
    \caption{Comparison of the mode I stress intensity factor $K_I$ for the SENT case: (a) comparison between the analytical solution and the vKMINN prediction for different crack length ratios $a/W$, and (b) effect of the I-integral radius on the extracted $K_I$ values for vKMINN and KMINN.}
    \label{fig:SENT_SIF_comparison}
\end{figure}

Fig.~\ref{fig:SENT_SIF_comparison}(a) shows the comparison of the mode I SIF $K_I$ between the analytical solution and the vKMINN prediction for different crack lengths $a$. The vKMINN results agree well with both the analytical solution and the KMINN framework, and all deviations remain within engineering acceptable accuracy of 3\%. Fig.~\ref{fig:SENT_SIF_comparison}(b) further examines the effect of the I-integral radius on the extracted SIFs for vKMINN and KMINN. Although the I-integral method is theoretically path-independent in LEFM, some radius dependence is still observed in the present implementation for KMINN. This is mainly because only the first-order Williams enrichment is adopted in KMINN method, while higher-order crack tip terms are neglected~\cite{Zhou.2026b}. As a result, the dominant near-tip singularity is captured well for KMINN, but the finite radius contour is still influenced by the omitted higher-order contributions. In addition, numerical discretization and finite radius integration introduce further sensitivity. In contrast, the outputs of vKMINN have direct relation to the singularity at crack tip. It does not rely on the high-order crack tip terms so the SIF evaluation is more stable and accurate than that of KMINN over a range of integration radii. This suggests that the crack field representation in vKMINN is more robust because the crack discontinuity and crack tip singularity are embedded explicitly, together with the holomorphic approximation of the regular field components $F_1$ and $F_2$.

\subsubsection{Double Edge Notch Tension}
\label{sec:double_edge_notch_tension}

\begin{figure}[!htbp]
    \centering
    \includegraphics[width=\textwidth,height=0.72\textheight,keepaspectratio]{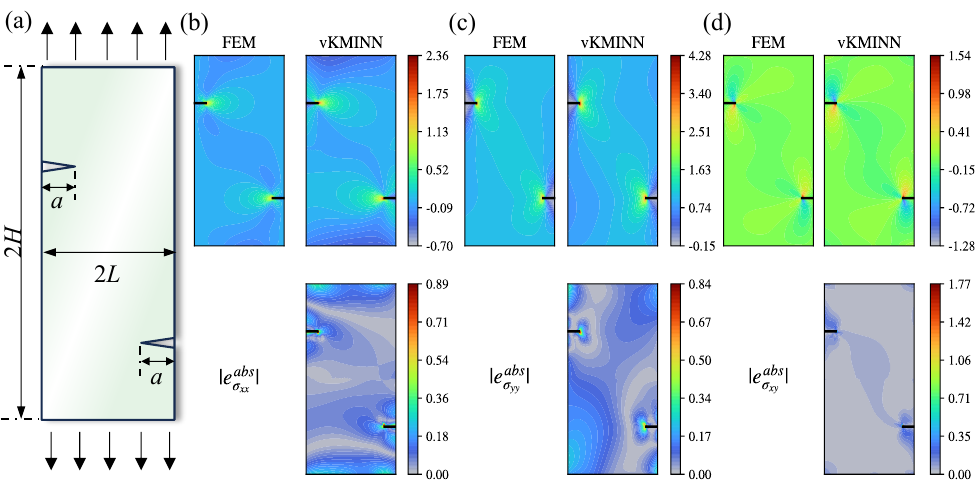}
    \caption{Analysis of the DENT case: (a) geometry and loading conditions, and comparison of FEM and vKMINN results with the corresponding absolute error fields for the stress components (b) $\sigma_{xx}$, (c) $\sigma_{yy}$, and (d) $\sigma_{xy}$.}
    \label{fig:double_edge_notch_tension}
\end{figure}

The second crack problem investigates the performance of the vKMINN framework for a double edge notch tension problem. This case is introduced to further assess the applicability of the present formulation beyond the single straight crack configuration discussed above. The geometry of the specimen is shown in Fig.~\ref{fig:double_edge_notch_tension} (a). The plate width is $2L=8$ mm, with $L=4$ mm, and the height is $2H=16$ mm. Two edge cracks of length $a=1$ mm are located on the opposite sides of the plate, and the vertical offset between the two crack tips is $8$ mm. A uniform tensile load of $\sigma_0=1$ MPa is applied on the external boundary, and the lower boundary is fixed. The material properties are $E=210$ GPa and $\nu=0.3$.

Fig.~\ref{fig:double_edge_notch_tension} compares the three stress components for the DENT case calculated by vKMINN with the FEM results. The proposed framework reproduces the global response of the specimen well and captures the near-tip fields of both cracks with good agreement. The corresponding mode I and mode II SIFs extracted at the two crack tips are summarized in Table~\ref{tbl:double_edge_notch_tension_sif}. The FEM reference gives $K_{\mathrm{I}}=2.62$ MPa$\sqrt{\mathrm{mm}}$ and $K_{\mathrm{II}}=-0.05$ MPa$\sqrt{\mathrm{mm}}$ for this configuration. The vKMINN predictions are $K_{\mathrm{I}}=2.51$ MPa$\sqrt{\mathrm{mm}}$ and $K_{\mathrm{II}}=-0.04$ MPa$\sqrt{\mathrm{mm}}$ for the left crack tip, and $K_{\mathrm{I}}=2.60$ MPa$\sqrt{\mathrm{mm}}$ and $K_{\mathrm{II}}=-0.05$ MPa$\sqrt{\mathrm{mm}}$ for the right crack tip. These results indicate that the proposed framework remains accurate for this double edge notch tension configuration, and provides reliable SIF evaluation for both crack tips.

\begin{table}[!htbp]
    \centering
    \caption{Stress intensity factors of the double edge notch tension case for FEM and vKMINN (unit: MPa$\sqrt{\mathrm{mm}}$).}
    \label{tbl:double_edge_notch_tension_sif}
    \begin{tabular}{lcccc}
        \toprule
        Crack tip & \multicolumn{2}{c}{FEM} & \multicolumn{2}{c}{vKMINN} \\
        \cmidrule(lr){2-3} \cmidrule(lr){4-5}
        & $K_{\mathrm{I}}$ & $K_{\mathrm{II}}$ & $K_{\mathrm{I}}$ & $K_{\mathrm{II}}$ \\
        \midrule
        Left crack  & 2.62 & -0.05 & 2.51 & -0.04 \\
        Right crack & 2.62 & -0.05 & 2.60 & -0.05 \\
        \bottomrule
    \end{tabular}
\end{table}

\subsubsection{Non-uniform Tension for Oblique Center Crack Tension}
\label{sec:non-uniform_tension_for_oblique_center_crack_tensile}

\begin{figure}[!htbp]
    \centering
    \includegraphics[width=\textwidth,height=0.66\textheight,keepaspectratio]{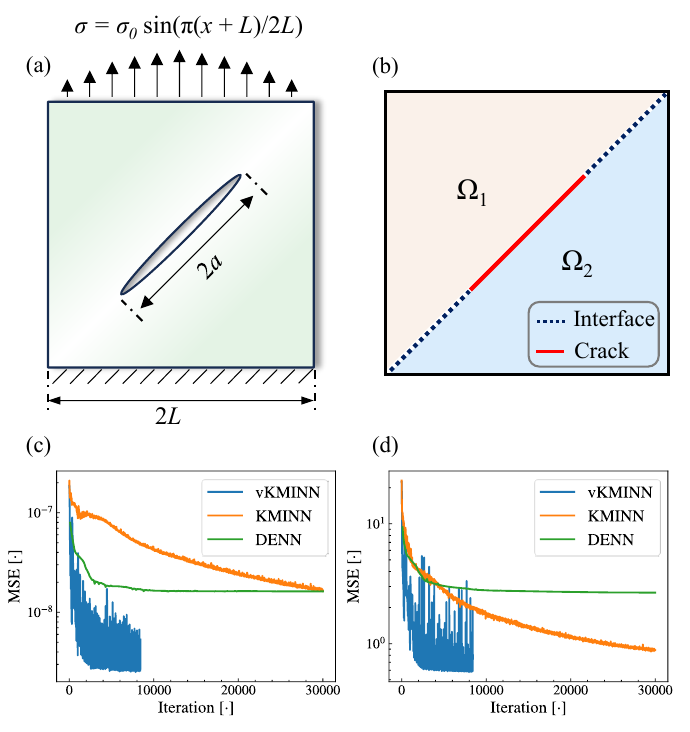}
    \caption{Geometry, loading conditions, domain decomposition, and training convergence for the OCCT case: (a) geometry and non-uniform loading conditions of the OCCT specimen, (b) subdomain decomposition and crack/interface definition used in KMINN, (c) convergence history of the MSE for the displacement fields, and (d) convergence history of the MSE for the stress fields.}
    \label{fig:OCCT_geometry_MSE_comparison}
\end{figure}

The third crack problem investigates the performance of the vKMINN framework for a mixed mode I/II crack problem in an OCCT specimen. The geometry of the OCCT specimen is shown in Fig.~\ref{fig:OCCT_geometry_MSE_comparison}(a). The square plate size is $2L = 8$ mm, and the crack half-length is $2a = 4$ mm. The crack is inclined at an angle of $\theta = 45^{\circ}$ with respect to the horizontal direction. A sinusoidal non-uniform tensile stress is applied on the top boundary as,

\begin{equation}
    \sigma_y = \sigma_0 \sin\left(\pi \dfrac{x+L}{2L}\right),
    \label{eq:non-uniform_tension_for_oblique_center_crack_tensile}
\end{equation}
where the maximum stress is $\sigma_0 = 10$ MPa at the center of the top boundary. The material properties are $E = 100$ GPa and $\nu = 0.3$.

DENN and KMINN frameworks are used to compare the accuracy and efficiency of the vKMINN framework. The subdomain decomposition used in KMINN is shown in Fig.~\ref{fig:OCCT_geometry_MSE_comparison}(b), whereas DENN and vKMINN do not require subdomains. All frameworks are trained using the same settings as those adopted for vKMINN in the SENT case. Fig.~\ref{fig:OCCT_geometry_MSE_comparison}(c) and (d) show the comparison of the mean squared error (MSE) during training for vKMINN, KMINN, and DENN. Table~\ref{tbl:occt_computational_cost} further summarizes the computational efficiency of the considered methods. Here, the MSE values are computed by pooling the three in-plane stress components and the two displacement components. The MSE is defined as,
\begin{equation}
    \text{MSE} = \frac{1}{N} \sum_{i=1}^{N} \left(\phi_i^{pred} - \phi_i^{ref}\right)^2,
\end{equation}
where $\phi_i^{pred}$ and $\phi_i^{ref}$ denote the predicted and reference values of a generic field component $\phi$ (stress or displacement) at the $i$-th pooled sample, with $\phi_i^{ref}$ taken from FEM, and $N$ is the total number of pooled samples. Although the training throughput of vKMINN is slightly lower than those of DENN and KMINN, the proposed framework reaches a stabilized solution with substantially fewer iterations than KMINN. As summarized in Table~\ref{tbl:occt_computational_cost}, vKMINN stabilizes after 8000 iterations, whereas DENN and KMINN require about 11000 and 30000 iterations, respectively. As a result, the wall-clock time of vKMINN remains markedly lower than that of KMINN and DENN. Moreover, the MSE of vKMINN remains consistently lower than those of KMINN and DENN, indicating that vKMINN achieves higher accuracy together with favorable computational efficiency in 206.93s wall-clock time for the present OCCT case.

\begin{table}[!htbp]
    \centering
    \caption{Comparison of computational efficiency for the OCCT case. For the neural network methods, the reported wall-clock time corresponds to the stabilization time. For FEM, the reported value is the total solution time.}
    \label{tbl:occt_computational_cost}
    \begin{tabular}{lccc}
        \toprule
        Model & Stable iteration & Iter./s & Wall-clock time (s) \\
        \midrule
        vKMINN & 8000  & 38.66 & 206.93 \\
        KMINN  & 30000 & 46.60 & 643.78 \\
        DENN   & 11000 & 45.22 & 243.26 \\
        FEM    & --    & --    & 10.20 \\
        \bottomrule
    \end{tabular}
\end{table}

Because no analytical solution is available for this problem, the FEM results are used as references to evaluate the SIFs $K_I$ and $K_{II}$ predicted by vKMINN, KMINN, and DENN. The SIFs are computed using the I-integral method with a radius of 0.5 mm. The results are shown in Table~\ref{tbl:non-uniform_tension_for_oblique_center_crack_tensile}. The vKMINN framework achieves the best accuracy, followed by KMINN and then DENN.
\begin{table}[!htbp]
    \centering
    \caption{Stress intensity factors of the OCCT case for FEM, vKMINN, KMINN, and DENN (unit: MPa$\sqrt{\mathrm{mm}}$).}
    \label{tbl:non-uniform_tension_for_oblique_center_crack_tensile}
    \begin{tabular}{llcccc}
        \toprule
        Crack tip                    & SIFs              & FEM   & vKMINN & KMINN & DENN  \\
        \midrule
        \multirow{2}{*}{Left crack}  & $K_{\mathrm{I}}$  & 12.49 & 12.29  & 11.00 & 12.22 \\
                                     & $K_{\mathrm{II}}$ & 13.50 & 13.60  & 13.87 & 13.71 \\
        \multirow{2}{*}{Right crack} & $K_{\mathrm{I}}$  & 20.26 & 20.21  & 18.13 & 18.54 \\
                                     & $K_{\mathrm{II}}$ & 14.64 & 14.79  & 14.43 & 12.74 \\
        \bottomrule
    \end{tabular}
\end{table}

\begin{figure}[!htbp]
    \centering
    \includegraphics[width=\textwidth,height=0.72\textheight,keepaspectratio]{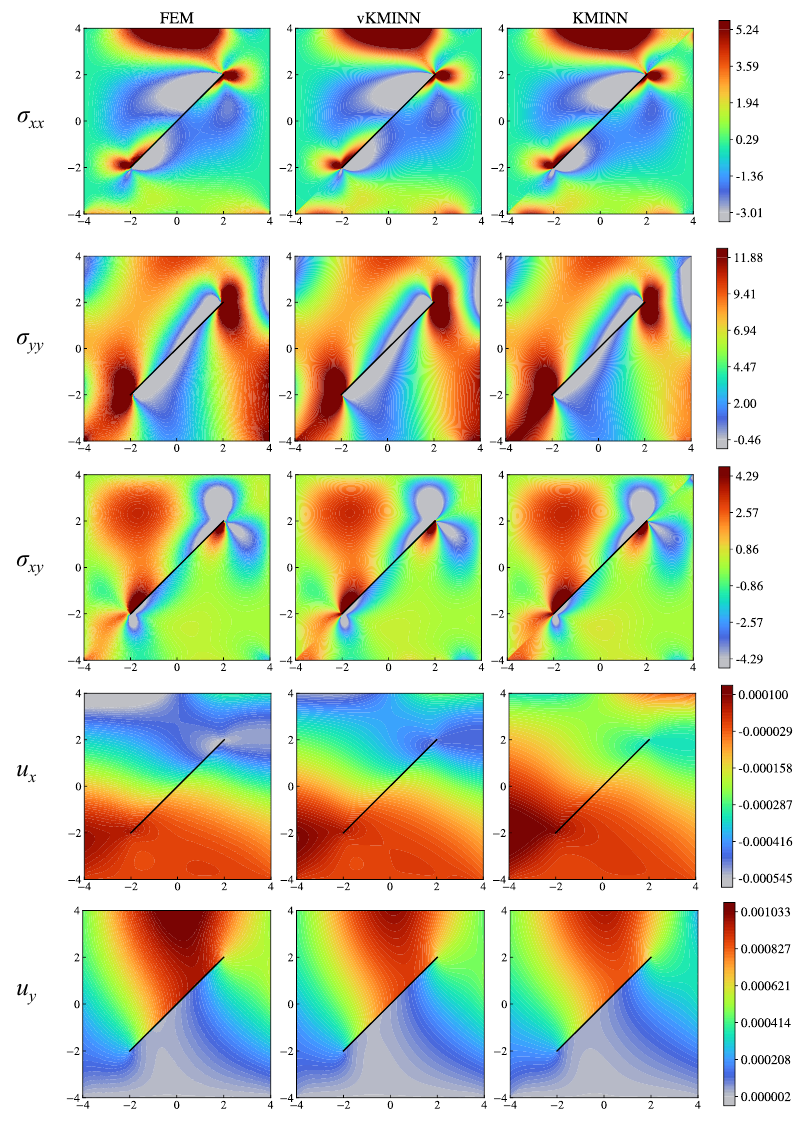}
    \caption{Comparison of the stress and displacement fields among FEM, vKMINN, and KMINN for the OCCT case: stress components $\sigma_{xx}$, $\sigma_{yy}$, and $\sigma_{xy}$, and displacement components $u_x$ and $u_y$.}
    \label{fig:OCCT_stress_and_displacement_comparison}
\end{figure}

Detailed comparisons of the stress and displacement fields among FEM, vKMINN, and KMINN are shown in Fig.~\ref{fig:OCCT_stress_and_displacement_comparison}. The stress fields predicted by vKMINN and KMINN agree well with the FEM results, while the displacement fields predicted by vKMINN show better agreement with FEM than those predicted by KMINN. These results indicate that the vKMINN framework can accurately and efficiently predict the stress and displacement fields near the crack tip.

\subsection{Discussion}
\label{sec:discussion}
The above results show that the proposed vKMINN framework provides an accurate and physically consistent variational framework for two-dimensional linear elastic problems with and without cracks. For the problems without cracks, vKMINN can accurately predict both stress and displacement fields under different loading conditions. Compared with the conventional PINN, the proposed method shows higher accuracy and better convergence. This result indicates the advantage of combining the KM representation with an energy-based training strategy, instead of learning the displacement field directly. For the mixed-boundary problem, the additional comparison with PIHNN shows that the improvement of vKMINN is not only caused by the holomorphic representation itself~\cite{Calafa.2024}. Since both PIHNN and vKMINN are based on the KM formulation, the observed difference highlights the effect of the training objective. At the same time, the results also show that PIHNN retains very high training efficiency in the crack-free case with pure displacement boundary conditions. Therefore, the main advantage of the proposed variational formulation in the present study lies not simply in faster training, but in providing a more accurate and more unified framework for the considered linear elastic and fracture problems.

For the crack problems, the comparison with KMINN shows that the present discontinuous potential representation is effective for embedding the crack face condition and the crack tip singularity directly into the solution ansatz \cite{Zhu.2025,Zhang.2026b,Zhao.2025}. As a result, vKMINN can provide accurate near-tip fields and reliable SIF predictions, while also stabilizing within the reported iteration range for the SENT, DENT, and OCCT cases. In the mixed mode OCCT case, the additional comparison with DENN further shows that the proposed framework is competitive not only against residual-based KM-informed methods, but also against an existing energy-based neural framework \cite{Zhao.2025}.

A possible reason for the improved solution quality and favorable convergence behavior of vKMINN is that the energy-based training objective is more consistent with the variational structure of linear elasticity within the KM framework. In the present formulation, the KM representation already satisfies the governing equations by construction. Therefore, the training process does not need to fit several governing residuals and boundary residuals separately. Instead, it only needs to search for the admissible holomorphic potentials that minimize the total potential energy. In this way, the optimization is guided by a single global mechanical objective. This reduces the competition among different loss terms and provides a more coherent training process \cite{Bischof.2025}. In addition, the total potential energy is an integral quantity, so it is generally less sensitive to local point oscillations than residual-based boundary losses \cite{Wang.2026}. As a result, the optimization process can be qualitatively understood as smoother and more stable for gradient-based training. Nevertheless, the results of the crack-free benchmark also show that the proposed variational formulation does not necessarily lead to faster training in every case. Therefore, the main advantage of vKMINN should be understood as the combination of improved accuracy, unified loss construction, and favorable convergence behavior in the considered benchmarks, rather than universally higher computational efficiency. A rigorous analysis of the loss landscape is beyond the scope of the present work.

Finally, the present study is restricted to two-dimensional isotropic linear elasticity. For three-dimensional problems, the current framework cannot be extended directly, because the KM representation is inherently a two-dimensional complex-potential theory. In addition, there are not analogous holomorphic two potentials formulation for general three-dimensional elasticity. Therefore, extending the present framework to three-dimensional problems would require a different analytical representation of the elastic field together with a new neural network structure and a corresponding variational formulation. In addition, the present framework is not directly applicable to nonlinear problems such as plasticity, damage, or large deformation. The current construction relies on the linear elastic KM theory and the associated crack tip asymptotics. Nevertheless, the proposed framework may still be useful as an efficient elastic predictor or pre-solution for nonlinear FEM analyses, where a sufficiently accurate elastic field could provide an improved initialization for subsequent incremental simulations \cite{Eshaghi.2026}. The proposed framework also shows potential for inverse problems in fracture mechanics. In principle, the holomorphic neural representation could be combined with sparse displacement or strain measurements to infer unknown quantities such as crack location or material parameters. Previous holomorphic neural network studies have already demonstrated the feasibility of crack location identification in this direction~\cite{Hund.2026}. However, several challenges remain, including the non-uniqueness of the inverse solution, the sensitivity to noisy and incomplete measurements, and the additional computational cost associated with repeated optimization. Moreover, the present work focuses on forward analysis. Thus, inverse identification and three-dimensional SIF evaluation are left for future investigation. Future work may extend the proposed framework to more general crack geometries, higher-order crack tip enrichments, and more complex material behaviors such as anisotropy or nonlinear constitutive models \cite{Manav.2024,Li.2025b}.

\section{Conclusions}
\label{sec:conclusions}
In this work, a variational Kolosov--Muskhelishvili informed neural network framework is proposed for solving two-dimensional linear elastic problems with and without cracks. By combining the complex variable formulation of linear elasticity with the principle of minimum total potential energy, the proposed method reformulates the mechanical problem as the minimization of an energy-based functional, while the stress and displacement fields are represented through two holomorphic KM potentials. For crack problems, a discontinuous stress-potential representation is further adopted to embed the discontinuity of the crack surface and the singularity of the crack tip directly into the solution ansatz.

The numerical results demonstrate that the proposed framework can accurately predict the stress and displacement fields for both regular elastic problems and fracture problems. For the no-crack cases, vKMINN shows consistently better accuracy than the baseline PINN under uniform and non-uniform boundary conditions. For the cracked cases, the framework provides accurate crack tip fields and reliable stress intensity factor predictions, showing good agreement with analytical solutions or FEM references. Compared with KMINN and DENN, the proposed vKMINN also exhibits a simpler loss construction and faster convergence in the considered cases.

Overall, the present study shows that the proposed vKMINN framework provides an effective and physically consistent variational approach for two-dimensional linear elastic fracture analysis. The results indicate that the combination of holomorphic KM potentials, energy minimization, and embedded crack representation offers a promising direction for improving both the efficiency and accuracy of solid mechanics solvers based on neural networks.

\section*{Acknowledgments}
The simulations were performed with computing resources granted by RWTH Aachen University, Germany, under project (rwth1654). The first author sincerely thanks the financial support of the China Scholarship Council (CSC: 202307000038).

\section*{Author Contributions}
Shuwei Zhou: Writing -- original draft, Conceptualization, Formal analysis, Visualization. Christian Häffner: Writing -- review \& editing, Formal analysis. Sophie Stebner: Writing -- review \& editing. Niklas Fehlemann: Writing -- review \& editing. Zhichao Wei: Writing -- review \& editing, Formal analysis. Sebastian Münstermann: Writing -- review \& editing, Supervision.

\section*{Data Availability}
The data that support the findings of this study are available from the corresponding author upon reasonable request.

\section*{Declaration of Competing Interest}
The authors declare that they have no known competing financial interests or personal relationships that could have appeared to influence the work reported in this paper.

\FloatBarrier
\renewcommand{\thefigure}{A.\arabic{figure}}
\renewcommand{\theHfigure}{A.\arabic{figure}}
\setcounter{figure}{0}
\section*{Analysis of the Penalty for the Dirichlet Boundary Condition}
\label{sec:analysis_of_the_penalty_for_the_dirichlet_boundary_condition}

The Dirichlet boundary condition in the proposed framework is imposed through the penalty term weighted by the factor $\alpha_u$. Since the choice of $\alpha_u$ affects the balance between the energy minimization term and the boundary-enforcement term, a sensitivity study is carried out here using the SENT benchmark introduced in Section~\ref{sec:crack_problems}. The same network architecture, sampling strategy, and training settings are used as those in the main SENT analysis, while only $\alpha_u$ is varied from $10^{-2}$ to $10^{5}$.

\begin{figure}[!htbp]
    \centering
    \includegraphics[width=\textwidth,height=0.72\textheight,keepaspectratio]{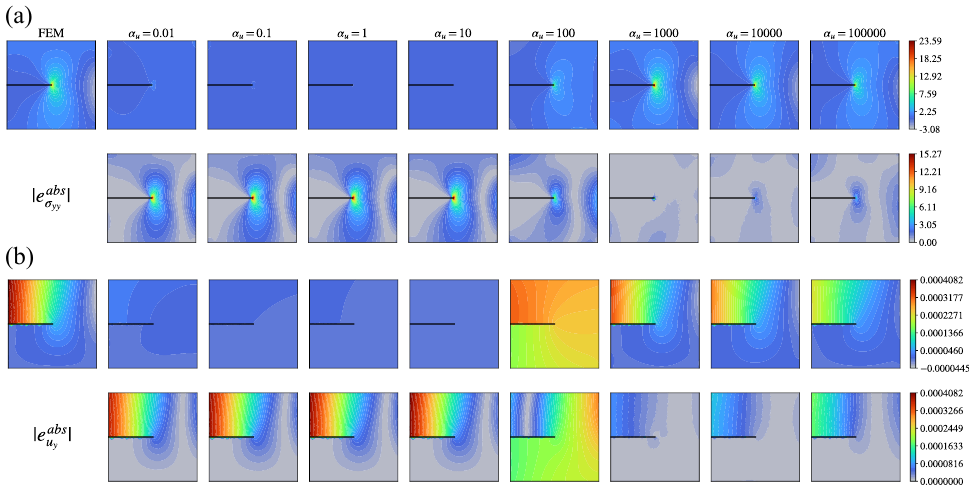}
    \caption{Comparison of the predicted fields and their absolute errors for different values of the Dirichlet penalty factor $\alpha_u$ in the SENT case: (a) stress component $\sigma_{yy}$ and (b) displacement component $u_y$.}
    \label{fig:alpha_u_sensitivity}
\end{figure}

Fig.~\ref{fig:alpha_u_sensitivity}(a) and (b) compare the predicted $\sigma_{yy}$ and $u_y$ fields and the corresponding absolute errors for different values of $\alpha_u$. The results show that the choice of $\alpha_u$ has a pronounced influence on the solution quality. When $\alpha_u$ is too small, the Dirichlet boundary condition is not enforced sufficiently, which leads to poor displacement predictions and inaccurate near-tip fields. As $\alpha_u$ increases to $1000$, the predicted fields improve significantly. However, when $\alpha_u$ becomes excessively large, the optimization is dominated by the boundary penalty term, which again deteriorates the prediction accuracy.

\begin{figure}[!htbp]
    \centering
    \includegraphics{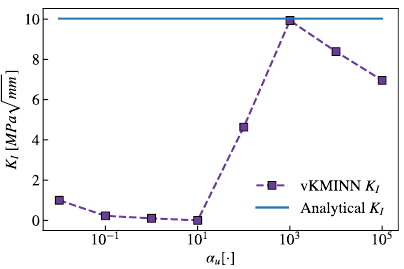}
    \caption{Comparison of the extracted mode-I stress intensity factor $K_I$ under different Dirichlet penalty factors $\alpha_u$ for the SENT case.}
    \label{fig:SIF_sensitivity_to_alpha_u}
\end{figure}

Fig.~\ref{fig:SIF_sensitivity_to_alpha_u} further shows the extracted mode-I SIF $K_I$ as a function of $\alpha_u$. The predicted $K_I$ is also highly sensitive to the choice of $\alpha_u$. Among the tested values, $\alpha_u=1000$ gives the best agreement with the analytical SIF result in the present SENT case and is therefore adopted in the main manuscript. These results indicate that the penalty factor should be chosen within a suitable intermediate range to balance boundary enforcement and energy minimization.

\FloatBarrier
\bibliographystyle{elsarticle-num}

\end{document}